\documentclass[preprint,times]{elsarticle}
\pdfoutput = 1
\usepackage{amsmath}  
\usepackage{amssymb}
\usepackage{subfigure}
\usepackage{ecrc}
\usepackage{multirow}
\usepackage{rotating}
\usepackage{xcolor}
\usepackage{listings}
\usepackage{latexsym}
\usepackage{bm}
\usepackage{bbm}
\usepackage{floatrow}
\usepackage{breqn}
\newfloatcommand{capbtabbox}{table}[][\FBwidth]
\usepackage{tikz}
\definecolor{mygreen}{RGB}{28,172,0} 
\definecolor{mylilas}{RGB}{170,55,241}

\volume{00}

\firstpage{1}
\runauth{}

\begin{document}

\begin{frontmatter}

\title{A size-dependent ductile fracture model: Constitutive equations, Numerical implementation and Validation}

\author[CEA,MINES]{J.M.~Scherer\corref{cor1}}
\author[CEA]{J.~Hure}

\cortext[cor1]{Corresponding author: jean-michel.scherer@cea.fr}
\address[CEA]{DEN-Service d'Etudes des Mat\'eriaux Irradi\'es, CEA, Universit\'e Paris-Saclay, F-91191, Gif-sur-Yvette, France}
\address[MINES]{MINES ParisTech, PSL Research University, MAT - Centre des mat\'eriaux, CNRS UMR 7633, BP 87 91003 Evry, France}

\begin{abstract}
  Size effects have been predicted at the micro- or nano-scale for porous ductile materials from Molecular Dynamics, Discrete Dislocation Dynamics and Continuum Mechanics numerical simulations, as a consequence of Geometrically Necessary Dislocations or due to the presence of a void matrix interface. As voids size decreases, higher stresses are needed to deform the material, for a given porosity. However, the majority of the homogenized models for porous materials used in ductile fracture modeling are size-independent, even though micrometric or nanometric voids are commonly observed in structural materials. Based on yield criteria proposed in the literature for nanoporous materials, a size-dependent homogenized model for porous materials is proposed for axisymmetric loading conditions, including void growth and coalescence as well as void shape effects. Numerical implementation of the constitutive equations is detailed. The homogenized model is validated through comparisons to porous unit cells finite element simulations that consider interfacial stresses, consistently with the model used for the derivation of the yield criteria, aiming at modeling an additional hardening at the void matrix interface. Potential improvements of the model are finally discussed with respect to the theoretical derivation of refined yield criteria and evolution laws.{\let\thefootnote\relax\footnote{{This article was presented at the IUTAM Symposium on Size-Effects in Microstructure and Damage Evolution at Technical University of Denmark, 2018.}}}
\end{abstract}


\end{frontmatter}

\section{Introduction}
Ductile fracture through void growth to coalescence has been extensively studied from experimental, theoretical and numerical perspectives \cite{besson2010,benzergaleblond,BLNT,pineaureview}, emphasizing the major roles played by porosity (void volume fraction) and stress triaxiality (ratio of the mean stress to the equivalent - deviatoric - stress). Numerous homogenized yield criteria for porous materials have been derived theoretically and validated numerically, incorporating the effects of porosity and stress triaxiality \cite{gurson,rousselier,thomason85a}, void shapes \cite{gologanu} and matrix material anisotropy \cite{benzergaanisotrope} or both \cite{monchiet,morinellipse,danas}. Complete sets of constitutive equations for porous materials have subsequently been proposed based on these yield criteria, requiring adding evolution laws for the additional internal state variables (porosity, void shape, ...), strain-hardening, multi-criteria selection (growth \textit{vs.} coalescence), as well as a finite-strain framework (see, \textit{e.g.}, \cite{benzergaleblond}). \textcolor{black}{Recent contributions tackle modelling of crystallographic effects observed when void size is below the grain size, \textit{i.e.}, porous single crystals, motivated by the numerous submicrometric voids revealed through X-ray tomography in standard structural steels \cite{daly} or by the use of single crystals in structural components. A strong effect of crystallographic orientations on void growth has been observed experimentally \cite{crepin,gan2} and numerically through 3D unit cell simulations \cite{liu,ha,yerra}. Homogenized yield criteria for porous single crystals have been developed to account for the effects of crystal orientation \cite{xuhan,paux,mbiakop,song1,hure2019}.}

Most of these homogenized models for porous materials are size-independent, assuming implicitly that only void volume fraction matters irrespectively of void size, \textcolor{black}{although porous materials with voids ranging from micrometric \cite{daly} down to nanometric sizes \cite{cawthorne,FISH1973} are encountered in industrial applications.} \textcolor{black}{Moreover, size effects have been predicted from theoretical and numerical studies}.
A \textcolor{black}{first} kind of size effect, occurring when voids size is lower than the dislocation mean free-path, has been revealed through Discrete Dislocation Dynamics (DDD) simulations \cite{segurado2009}. In such situations, dislocations exhaustion can lead to the absence of void growth under mechanical loading. A \textcolor{black}{second} kind of size effect is related, in a broad sense, to an additional hardening at or close to the void matrix interface. Strain gradient (crystal-)plasticity models - accounting for the presence of Geometrically Necessary Dislocations \cite{ashby} to extend conventional plasticity to lower scales - have been used in porous unit cells simulations (see \cite{fleckhutchinson2001,niordson2008,borg2008,chaothese} and reference therein), showing a strong effect of the void size \textcolor{black}{on both void growth rate and } strength of the porous material, consistently with DDD simulations \cite{hussein}. Numerous Molecular Dynamics (MD) studies have also been performed to assess the strength of (nano-)porous materials, considering voids in an initially dislocation free matrix material (\cite{traiviratana,mi,brach} and references therein). Plasticity occurs through dislocation emission from void matrix interface \cite{lubarda}, and an influence of surface tension on dislocation emission has been observed \cite{wilkerson,chang2013}, leading to higher strength for smaller voids. \textcolor{black}{While theoretical arguments and numerical predictions regarding void size effects appear sound, it should be emphasized that, to the authors' knowledge, no clear experimental evidence of void size effect on growth rate has been reported so far.}

Size-dependent homogenized yield criteria for porous materials have been proposed incorporating such additional hardening at or close to the void matrix interface. Considering strain-gradient plasticity for the matrix material, isotropic Gurson-type (assuming spherical voids) yield criteria have been derived in \cite{wen2005,li2006,monchiet2013b} that depend on the ratio between the void size and the lengthscale introduced through the strain-gradient plasticity model. Adding evolution laws, the homogenized model for porous materials based on the criterion proposed in \cite{monchiet2013b} shows hardening as voids size decreases. Considering interface stresses, size-dependent isotropic yield criteria for both spherical and spheroidal voids in the growth regime have been derived in \cite{dormieux2010,monchiet2013} and validated in \cite{morinthese,morin2015}, the latter describing a homogenized model based on the yield criterion for spherical voids \cite{dormieux2010}. A coalescence criterion has been proposed considering interface stresses, for spheroidal voids \cite{gallican}. \textcolor{black}{All these yield criteria for porous materials were derived in the continuum mechanics framework, assuming that plastic flow occurs at a scale well below the void size, which might be questionable for very small voids. Recent experimental observations however show clear evidences of homogeneous deformation of nanometric voids \cite{margolin2016,ding2016}, justifying the use of continuum models for porous materials down to the nanometric range, at least for applications involving large strain levels and/or large initial dislocations (sources) density}. To the authors' knowledge, no size-dependent homogenized yield criterion for porous single crystals is available in the literature. In addition, no size-dependent homogenized model for porous materials including void growth and coalescence regimes, as well as voids shape effects, has been described in the literature.\\

Therefore, the aim of this study is to provide a size- and shape-dependent homogenized model for isotropic porous materials, with a description of the numerical implementation as well as validation of the model through comparisons to unit cells simulations. Yield criteria proposed in \cite{monchiet2013} for the growth regime and \cite{gallican} for the coalescence regime that consider spheroidal (nano-)voids are used. In section~2, the constitutive equations are detailed, as well as the numerical implementation. Section~3 describes the reference finite strain unit cells simulations and the comparisons between the unit cells results and homogenized model predictions. The perspectives of this study are finally discussed in Section~4.

\section{Description of the size-dependent homogenized model for isotropic porous materials}

Underline $\underline{A}$ and bold $\textbf{A}$ symbols refer to vectors and second-order tensors, respectively. A cartesian orthonormal basis $\{\underline{e}_x,\underline{e}_y,\underline{e}_z\}$  along with coordinates $\{x,y,z \}$ are used. $\textbf{A}'$ is the deviator of a second-order tensor $\textbf{A}' = \textbf{A} - [\text{tr}\textbf{A}/3]\, \textbf{I}$. Equivalent stress and strain-(rate) are computed according to $\sigma_{eq} = \sqrt{[3/2]\bm{\sigma'}:\bm{\sigma'}}$ and $\dot{\varepsilon}_{eq} = \sqrt{[2/3]\bm{\dot{\varepsilon}'}:\bm{\dot{\varepsilon}'}}$.

\subsection{Synthesis of the modeling hypothesis}

The starting point of the size-dependent yield criteria for porous materials derived in \cite{monchiet2013} and \cite{gallican} is the concept of interface stresses developed by Gurtin \& Murdoch \cite{gurtin}. Dormieux \& Kondo \cite{dormieux2010} proposed, for nanoporous materials, to consider a void matrix interface with continuity of the displacement and discontinuity of the traction vector, through the presence of either residual or deformation interface stresses. In the framework of limit analysis \cite{benzergaleblond} which deals only with finding the strength of some Representative Volume Element (RVE), and restricting to tensile loading, both approaches (residual \textit{vs.} deformation interfacial stresses) lead to similar predictions regarding the strength of porous materials. From an energetic point of view, such modeling leads to an interfacial dissipation $d\mathcal{E}_{int} \sim \gamma d\mathcal{S}$ (with $\gamma$ the surface energy) to be compared to the dissipation of the matrix material around voids. As a working example, for the hydrostatic loading of a porous isotropic material, with spherical voids of radius $R$, the interfacial dissipation can be written as $d\mathcal{E}_{int} \sim \gamma R dR$ while the plastic dissipation in the matrix material can be written as $d\mathcal{E}_{mat} \sim \sigma_0 R^2 dR$ (with $\sigma_0$ the matrix material yield stress). Balancing both equations leads to the definition of a natural dimensionless strength of the interface:
\begin{equation}
  \Gamma = \frac{\gamma}{\sigma_0 R}
  \label{defgamma}
\end{equation}
that relates the interfacial dissipation to the volumic dissipation. For $\Gamma \ll 1$, the strength of the interface is negligible compared to the strength of the bulk that sets the strength of the porous materials. On the contrary, for $\color{black}\Gamma \sim 1 \color{black}$, the strength of the RVE will strongly depend on the interface strength, leading to size-dependent yield criteria as shown by the inverse dependence of $\Gamma$ on the void size in Eq.~\ref{defgamma}. Following this approach and the yield criteria derived in \cite{monchiet2013} and \cite{gallican}, in order to provide a size-dependent homogenized model for porous isotropic materials, the next step forward is the numerical implementation of the model and its validation. This requires adding elasticity, hardening and evolution laws for the state variables to the yield criteria, performing the numerical integration of these constitutive equations, and finally validating the homogenized model with respect to porous unit cell simulations with the same modeling hypothesis than for the theoretical derivation. Such step has already been tackled partially in \cite{morin2015} with the numerical implementation of growth yield criterion for spherical voids \cite{dormieux2010}, without validation through comparisons to unit cell simulations. Therefore, in the following, spheroidal voids are considered, for both growth and coalescence regimes.

\subsection{Constitutive equations}
\label{consti}
\begin{figure}[H]
\centering
\includegraphics[height = 5cm]{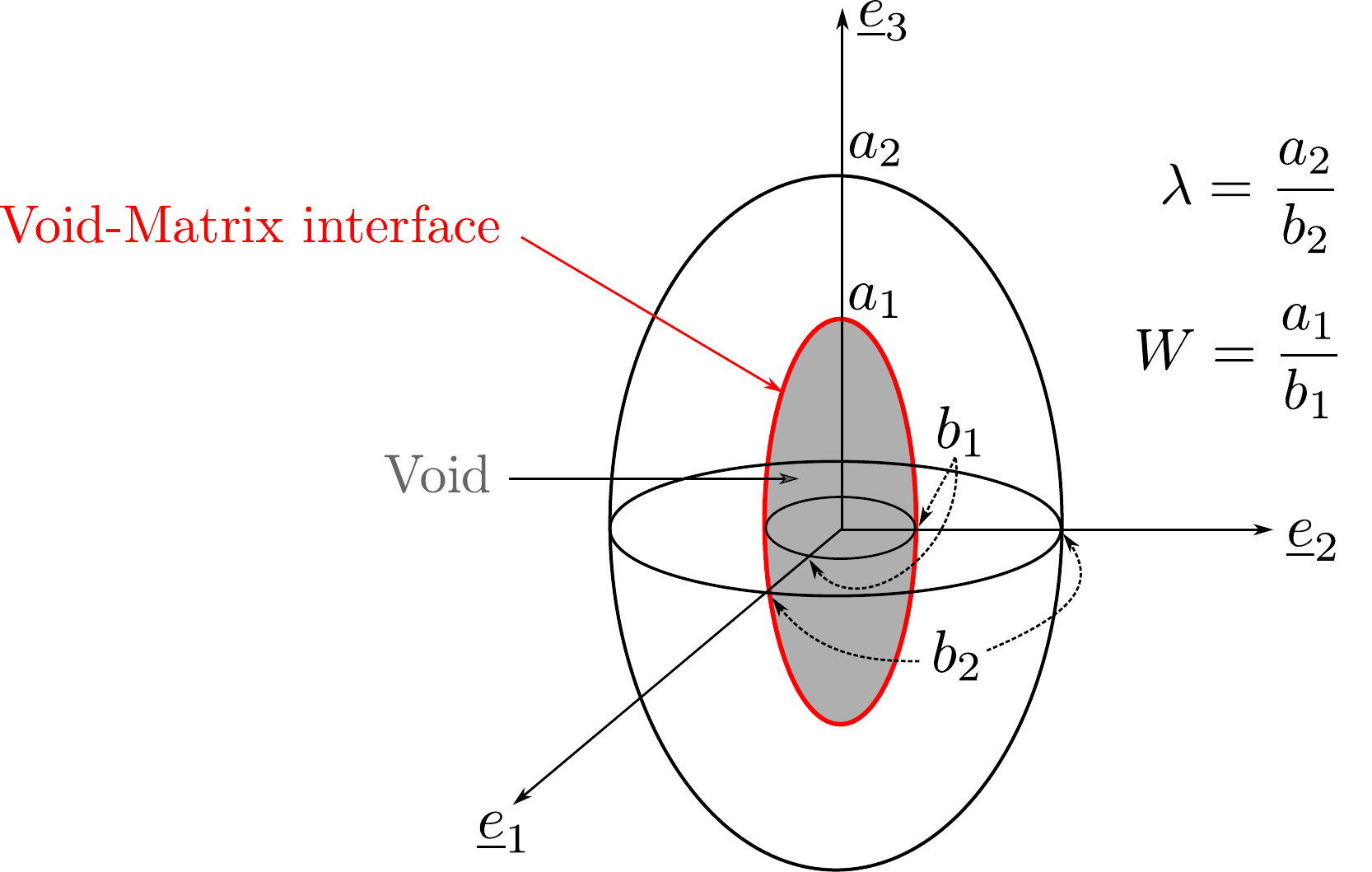}
\caption{Unit cell used for the derivation of the yield criterion in \cite{monchiet2013}, considering a spheroidal void in a confocal spheroidal unit cell. Definition of the geometrical parameters: void $W$ and cell $\lambda$ aspect ratios}
\label{figgrowth}
\end{figure}

The key ingredients of the size-dependent homogenized model for isotropic porous materials are the yield criteria derived in \cite{monchiet2013} in the growth regime, \textit{i.e.} without strong interactions between voids, and in \cite{gallican} for the coalescence regime, \textit{i.e.} when adjacent voids strongly interact \cite{koplik}. A yield criterion in the growth regime has been derived in \cite{monchiet2013} considering a matrix material surrounding voids obeying von Mises perfect plasticity with yield stress $\sigma_0$ and a void matrix interface obeying (2D) von Mises plasticity with (2D) yield stress $\gamma$, as a modelling of interface stresses. \color{black}Spheroidal voids, with semi-axis $a_1$, $b_1$ and $b_1$, were considered in a spheroidal confocal unit cell (Fig.~1), with semi-axis $a_2$, $b_2$ and $b_2$. Under axisymmetric loading conditions where the main loading direction is oriented along $\underline{e}_3$ (Fig.~\ref{figgrowth}) porosity $f$, void aspect ratio $W$, cell aspect ratio $\lambda$ and intervoid distance $\chi$ are respectively defined as:
\begin{equation}
	\begin{aligned}
		f = \frac{a_1b_1^2}{a_2b_2^2}, \ \ \ \ \
		W = \frac{a_1}{b_1}, \ \ \ \ \
		\lambda = \frac{a_2}{b_2}, \ \ \ \ \
		\chi = \frac{b_1}{b_2}
	\end{aligned}
	\label{param}
\end{equation}
\color{black}
 A dimensionless interfacial strength can be defined as:
\begin{equation}
  \Gamma = \frac{\gamma}{\sigma_0 a_1}
  \label{eqgamma}
\end{equation}
where the material lenthscale $\gamma/\sigma_0$ is compared to the revolution axis length of the spheroid $a_1$\footnote{Another (arbitrary) choice would have been to use the axis $b_1$.}. The yield criterion can not be recasted into a closed form expression, and the parametric form for axisymmetric loading conditions is:
\begin{equation}
\Phi_{\mathrm{Growth}}^{\mathrm{spheroidal}}\leq 0 \left\{\begin{aligned}
\frac{\sigma_m}{\sigma_0}=  \mathcal{G}_1(\Gamma,\bm{\alpha},\xi) &=\frac{\mathcal{U}(\xi)}{f\kappa}\left[1+\frac{\eta}{3}(1-3\alpha_2)\right] -\frac{1}{3}(1-\zeta)\mathcal{V}(\xi)\left[\xi+(1-3\alpha_2)\right](1-3\alpha_2)\\
&+\Gamma\frac{a_1S_1}{3V_1}\frac{h_1+h_3\xi}{\mathcal{Z}(\xi)}\\
\frac{\sigma_{eq}}{\sigma_0} = \mathcal{G}_2(\Gamma,\bm{\alpha},\xi)&=\frac{\eta\mathcal{U}(\xi)}{f\kappa}-(1-\zeta)\mathcal{V}(\xi)\left[\xi+(1-3\alpha_2)\right]\\
&+ \Gamma\frac{a_1 S_1}{V_1}\frac{h_2\xi+h_3}{\mathcal{Z}(\xi)}\\
\end{aligned} \right.
\label{critgrowth1}
\end{equation}
where $\sigma_{eq}$ is the von Mises equivalent stress, $\sigma_m = \sigma_{kk}/3$ the mean stress, $f$ the porosity, $\sigma_0$ and $\gamma$ the yield stresses of the matrix material and interface, respectively. The yield criterion also depends on a list of geometrical parameters \cite{monchiet2013} that are defined in Appendix A:
\begin{equation}
  \bm{\alpha} = \{ \mathcal{U},\mathcal{V},\mathcal{Z},\kappa,\eta,\zeta,\alpha_2,b_1,S_1,V_1,h_1,h_2,h_3 \}
  \label{eqalpha}
\end{equation}
\noindent
\textcolor{black}{In case of spherical voids ($W=1$) without size effects ($\Gamma=0$) the yield criterion described by Eq.~\ref{critgrowth1} reduces to Gurson's criterion. A closed-form expression of the yield criterion described by Eq.~\ref{critgrowth1} can also be written without size effects, that requires however an additional parameter to lead to good predictions when compared to porous unit cell results considering spheroidal voids in orthorhombic cells \cite{pardoen} (as the one considered in Section~3.1 for unit cells simulations).}
Unfortunately, due to the absence of closed-form expression for Eq.~\ref{critgrowth1} in the general case, this additional parameter is introduced in a different way where the porosity is replaced by an effective porosity:
\begin{equation}
  q_W f \rightarrow f
  \label{qW}
\end{equation}
which is exactly equivalent to \cite{pardoen} for spherical and prolate voids. The parameter $q_W$ calibrated in \cite{pardoen} for $\Gamma = 0$ is used in this study (see Appendix~A for the mathematical expression), and an extension is proposed for $\Gamma \neq 0$ in Section~3.\\

\noindent
\color{black}Coalescence stresses for porous materials accounting for interfacial stresses have been derived in \cite{gallican}, considering cylindrical voids of radius $b_1$ and height $a_1$ in cylindrical unit cells of radius $b_2$ and height $a_2$. In that case although the void and cell shapes are different from the one used for the growth criterion, definitions of aspect ratio $W$ and intervoid distance $\lambda$ in Eq.~\ref{param} are kept identical. \color{black} Such coalescence stress - corresponding to the macroscopic stress perpendicular to the coalescence plane - can be rewritten as a yield criterion \cite{benzergaleblond} - involving only invariants of the stress tensor - which takes the closed form expression\footnote{The term proportional to $\Gamma$ in Eq.~\ref{critcoa} differs from the one derived in \cite{gallican} due the definition of $W$ (with respect to $b_1$ in \cite{gallican} instead of $a_1$ in this study).}:
\begin{equation}
  \Phi_{\mathrm{Coalescence}}^{\mathrm{spheroidal}} =  \frac{\sigma_{eq}}{\sigma_0}+\frac{3}{2}\frac{\sigma_m}{\sigma_0}-\frac{3}{2}C_f(\chi,W) - \frac{3\Gamma}{\sqrt{3}}W\sqrt{1+3\chi^4} \leq 0
  \label{critcoa}
\end{equation}
where $C_f$ is the coalescence stress without size effects ($\Gamma = 0$).
Different formula have been proposed in the literature for the parameter $C_f$ based on limit analysis and heuristic corrections. The expression proposed in \cite{torki} is used:
\begin{equation}
 C_f(\chi,W) = t(W,\chi) \left[\frac{\chi^3 - 3\chi + 2}{3\sqrt{3} W\chi} \right]     + \frac{b}{\sqrt{3}}\left[2 - \sqrt{1+3\chi^4} + \ln \frac{1 + \sqrt{1 + 3\chi^4}}{3\chi^2}    \right]
\label{eqcf}
\end{equation}
with $b=0.9$ and $t(W,\chi) = [W(-0.84+12.9\chi)]/[1+W(-0.84+12.9\chi)]$. Moreover, Eqs.~\ref{critcoa},~\ref{eqcf} have been derived for a hexagonal lattice of cylindrical voids (through the approximation of considering a cylindrical unit-cell). For spheroidal voids in orthorhombic cells (as the one considered in Section~3.1 for unit cells simulations), an effective intervoid distance can be considered:
\begin{equation}
  q_{\chi}\chi \rightarrow \chi
  \label{qchi}
\end{equation}
following the proposition of \cite{torki} that is based on considering an effective porosity in the coalescence layer. As for the growth criterion where a parameter $q_{W}$ is used (Eq.~\ref{qW}), the parameter $q_{\chi}$ aims at correcting the yield criterion derived for a simplified unit cell - to make analytical derivation possible - to another - more realistic - one. While the yield criteria in growth and coalescence regimes (Eqs.~\ref{critgrowth1},~\ref{critcoa}) are the main ingredients of the model, they shall be supplemented by a description of elasticity as well as with evolutions laws for the internal state variables. Elasticity is assumed to obey Hooke's law:
\color{black}
\begin{equation}
  \pmb{\sigma} = \frac{E}{1+\nu}\left(\pmb{\varepsilon}_e+\frac{\nu}{1-2\nu}\text{tr}(\pmb{\varepsilon}_e)\mathbf{I}\right)
  \end{equation}
where $E$ is Young's modulus and $\nu$ the Poisson's ratio. \color{black} Additive decomposition of elastic and plastic strains is used:
\begin{equation}
  \pmb{\varepsilon} = \pmb{\varepsilon}_e+\pmb{\varepsilon}_p
  \label{partition}
\end{equation}
\textcolor{black}{At a given porosity,} limit-analysis framework used to derive the yield criteria keep the normality rule at the macroscale:
\begin{equation}
  \dot{\pmb{\varepsilon}}_p=\dot{\Lambda}\displaystyle{\frac{\partial \Phi}{\partial \pmb{\sigma}}} \ \ \ \ \ \ \ \ \ \ \dot{\Lambda}\Phi(\pmb{\sigma})=0 \text{ et } \dot{\Lambda} \geq 0
  \label{epp}
\end{equation}
Evolution laws for the geometrical parameters are also required, and are not provided by limit-analysis. Volume conservation leads to the classical evolution for the porosity:
\begin{equation}
  \dot{f}=(1-f)\dot{\varepsilon}_{p,kk}
  \label{evolporo}
\end{equation}
The evolution of all other geometrical parameters - noted $\bm{\alpha}$ (Eq.~\ref{eqalpha}, Appendix~A) - are related to the evolution of the void aspect ratio $W$ and the unit cell aspect ratio $\lambda$, that depends on the active deformation mode. In the growth regime, the evolutions of unit cell and void aspect ratio are:
\color{black}
\begin{equation}
  \mathrm{Growth\ regime}\ \ \ \ \ \ \ \ \ \ \left\{  \begin{aligned}
\textcolor{black}{\dot{\lambda}} &= \textcolor{black}{\lambda \left( \underline{e}_V \dot{\bm{\varepsilon}}_{p} \underline{e}_V - \underline{e}_T \dot{\bm{\varepsilon}}_{p}\underline{e}_T    \right)} \\
\textcolor{black}{\frac{\dot{W}}{W}} &=  \textcolor{black}{\mathcal{W}(T,\Gamma)\, \underline{e}_V \dot{\bm{\varepsilon}}_{p}^' \underline{e}_V   + \left[\frac{1-3\beta_1}{f}  + 3 \beta_2 - 1    \right] \textbf{I}:\dot{\bm{\varepsilon}}_p}
  \end{aligned} \right.
  \label{evolgrowth}
\end{equation}
where expressions of $\beta_1$ and $\beta_2$ are given in Appendix~A, \textcolor{black}{with $\underline{e}_V$ the void axis (with respect to which the parameter $W$ is defined, \textit{e.g.}, $\underline{e}_3$ in Fig.~1) and $\underline{e}_T$ an axis perpendicular to $\underline{e}_V$. Evolution law for the void axis $\underline{e}_V$ can be used as discussed in \cite{benzergaleblond}, but the complete model is mainly restricted to axisymmetric loading conditions with respect to the initial void axis, due to the yield criteria used.} \color{black} The first equation of Eq.~\ref{evolgrowth} comes from geometrical consideration, and corresponds to the elongation (or contraction) of the porous unit cell, under the assumption of axisymmetric loading conditions. The form of the evolution of the void aspect ratio is the same as the ones used in \cite{benzergaleblond}. The function $\mathcal{W}$ is an heuristic correction that should be calibrated against comparisons to numerical unit cells computations. Different expressions are available in \cite{benzergaleblond} and \cite{pardoen} for $\mathcal{W}$ but only for the case $\Gamma = 0$. An expression for $\mathcal{W}$ has been calibrated in this study, and is given in Appendix~A. For completeness, it should be noticed that non-linear variational approaches have provided void aspect ratio evolution laws without resorting to heuristic corrections. However, the case of porous materials with void matrix interface stresses has not been considered to date.\\

\noindent
Similarly, Eq.~\ref{evolcoa} describes the evolutions of the unit cell aspect ratio $\lambda$, void aspect ratio $W$ in the coalescence regime that comes from both uniaxial straining conditions prevailing at coalescence and material incompressibility in the coalescence layer (of height assumed to be equal to the void height) \cite{benzergaleblond}:
\begin{equation}
\mathrm{Coalescence\ regime}\ \ \ \ \ \ \ \ \ \ \left\{  \begin{aligned}
\dot{\lambda} &= \frac{3}{2}\lambda \dot{{\varepsilon}}^p_{eq} \\
\dot{W} &= \frac{9}{4} \frac{\lambda}{\chi} \left[1 - \frac{2}{\pi \chi^2}    \right]\dot{{\varepsilon}}^p_{eq} 
\end{aligned} \right. 
\label{evolcoa}
\end{equation}
\color{black}
Transverse intervoid distance $\chi$ is also required for the coalescence criterion, and to that extent its evolution has also to be computed in the growth regime. Since $\chi$ does not enter in the growth criterion or growth evolution laws, its value is calculated with its definition for a spheroidal void in an orthorhombic unit cell which corresponds to the void arrangement considered in Section \ref{ucsimu}. It is then corrected by the previously mentionned parameter $q_{\chi}$ before entering the coalescence criterion which has been derived for a cylindrical porous unit cell.
Hence in both cases (growth and coalescence), the expression of the transverse intervoid distance in Eq.~\ref{param} can be rewritten as:
\begin{equation}
  \chi = \left[\frac{6f\lambda}{\pi W} \right]^{1/3}
  \label{eqchi}
\end{equation}
\color{black} The evolutions of void semi-axis $a_1$ and $b_1$ can be computed based on porosity and void aspect ratio evolutions, required to update the dimensionless interfacial strength $\Gamma$: 
\begin{equation}
  \frac{\dot{a}_1}{a_1} = \frac{1}{3}\left(\frac{\dot{f}}{f(1-f)} + 2\frac{\dot{W}}{W}    \right) \ \ \ \ \ \ \ \ \ \  \frac{\dot{b}_1}{b_1} = \frac{1}{3}\left(\frac{\dot{f}}{f(1-f)} - \frac{\dot{W}}{W}    \right)
  \end{equation}
\color{black}
A criterion for deformation mode (growth \textit{vs.} coalescence) selection is necessary. Inspired by the method proposed in \cite{besson2010} two equivalent stresses $\sigma_{\mathrm{Growth}}^*$ and $\sigma_{\mathrm{Coalescence}}^*$ (standing for $\sigma_0$ in the yield criteria) are implicitely defined such that:
\begin{equation}
	\Phi_{\mathrm{Growth}}^{\mathrm{spheroidal}}(\sigma_{\mathrm{Growth}}^*) = 0 \ \ \ \ \ \ \ \ \ \  	\Phi_{\mathrm{Coalescence}}^{\mathrm{spheroidal}}(\sigma_{\mathrm{Coalescence}}^*) =0 
\end{equation}
For the growth regime $\sigma_{\mathrm{Growth}}^*$ is calculated using a Newton-Raphson algorithm, while for the coalescence regime $\sigma_{\mathrm{Coalescence}}^*$ is obtained directly. 
The active deformation mode is the one for which the equivalent stress is the greatest. Therefore the implemented yield criterion selection is written as follows:
\begin{equation}
\Phi^{\mathrm{spheroidal}}\leq 0 \left\{\begin{aligned}
(\sigma_{\mathrm{Growth}}^* - \sigma_0) \leq 0 \ \ \ \text{if } (\sigma_{\mathrm{Growth}}^* \geq \sigma_{\mathrm{Coalescence}}^*)\\
(\sigma_{\mathrm{Coalescence}}^* - \sigma_0) \leq 0 \ \ \ \text{if } (\sigma_{\mathrm{Growth}}^* < \sigma_{\mathrm{Coalescence}}^*)
\end{aligned} \right.
\label{critgrowthselect}
\end{equation}
\color{black}
Finally, hardening of the matrix material is included following Gurson's proposal \cite{gurson} that assumes that the macroscopic plastic dissipation $\bm{\sigma}:\dot{\pmb{\varepsilon}}^p$, where $\bm{\sigma}$ and $\dot{\pmb{\varepsilon}}^p$ are the homogenized stress and plastic strain-rate, respectively, is equal to the sum of local plastic dissipation, assuming constant average plastic strain $p$ in the matrix material, leading to:
\color{black}
\begin{equation}
  (1 - f)\sigma_0(p)\dot{p} + \frac{2}{3}\frac{S_1}{V_1}\gamma \dot{\varepsilon}_{p,kk} = \bm{\sigma}:\dot{\pmb{\varepsilon}}^p
  \label{hardening}
  \end{equation}
where the first term of the left-hand side of Eq.~\ref{hardening} is the classical one, while the second term is introduced phenomenologically here to account for the contribution of the interface to the macroscopic plastic dissipation, with $S_1$ and $V_1$ the surface and volume of the void, respectively. In particular, for a spherical void of radius $R$ under hydrostatic loading, this additional term is consistent with the dissipated energy through surface extension of a sphere $\mathrm{d}\mathcal{E}_{int}=\gamma\mathrm{d}S_1$.
\color{black}
\subsection{Numerical implementation}

A fully implicit integration scheme for rate-independent plasticity models is often recommended in order to satisfy all equations at the end of the time step, ensuring numerical stability. However, the complexity of the governing equations for the geometrical parameters (Appendix~A) prevents the analytical calculation of the jacobian matrix (required to solve the set of non-linear equations obtained by discretizing implicitly the constitutive equations through Newton-Raphson method). Therefore, the constitutive equations described in Section~\ref{consti} are discretized according to an implicit scheme for the elastic strain tensor $\bm{\varepsilon}_e$, plastic multiplier $\Lambda$, void volume fraction $f$ and average plastic strain in the matrix material $p$. All other parameters are discretized explicitly, \textit{i.e.}, are fixed during the time step at their initial value and updated at the end of the time step.\\

\noindent
Plastic flow rule requires to compute the derivative of the yield criterion with respect to the stress tensor, which is straightforward for the coalescence criterion:
\begin{equation}
  \begin{aligned}
    \frac{\partial \Phi_{\mathrm{Coalescence}}}{\partial \bm{\sigma}} &= \frac{1}{\sigma_0} \left[\frac{\partial \sigma_{eq}}{\partial \bm{\sigma}} + \frac{3}{2} \frac{\partial \sigma_{m}}{\partial \bm{\sigma}} \right] \\
  &= \frac{1}{\sigma_0} \left[ \frac{3 \bm{\sigma}'}{2\sigma_{eq}} + \frac{\textbf{I}}{2}     \right]
\end{aligned}
\end{equation}
but more intricate when the yield criterion is only available in a parametric form, as for Eq.~\ref{critgrowth1}. However, Morin \textit{et al.} \cite{morin2015} provided an expression of the plastic normal direction:
\begin{equation}
  \frac{\partial\Phi_{\mathrm{Growth}}}{\partial\pmb{\sigma}}=C^2\left(-\frac{\partial\sigma_{eq}}{\partial\xi}\frac{\mathbf{I}}{3} + \frac{\partial\sigma_{m}}{\partial\xi}\frac{3}{2}\frac{\pmb{\sigma}'}{\sigma_{eq}} \right)
\end{equation}
For given initial values $\{\bm{\varepsilon}_e^n,p_n,f_n,\bm{\alpha}_n\}$ and total strain increment $\Delta \bm{\varepsilon}$, the implicit system of equations (corresponding to the discretization of Eq.~\ref{partition},~\ref{epp},~\ref{evolporo},~\ref{hardening}) with respect to $\{\Delta \bm{\varepsilon}_e,\Delta \Lambda,\Delta f,\Delta p \}$ to be solved is:
\begin{equation}
  \begin{aligned}
    \Delta\pmb{\varepsilon} - \Delta\pmb{\varepsilon}_e-\Delta\Lambda \displaystyle{\frac{\partial \Phi}{\partial \pmb{\sigma}}}\left(\bm{\sigma}_{n+1},p_{n+1},f_{n+1},\bm{\alpha}_n  \right) &=0\\
    \Delta\Lambda \Phi\left(\bm{\sigma}_{n+1},p_{n+1},f_{n+1},\bm{\alpha}_n  \right) &= 0\\
    \Delta f - (1-f_{n+1}) \Delta \Lambda \mathrm{tr}\left[\Phi\left(\bm{\sigma}_{n+1},p_{n+1},f_{n+1},\bm{\alpha}_n \right)\right]  &= 0\\
    \pmb{\sigma}_{n+1}\text{ : }\Delta\Lambda \displaystyle{\frac{\partial \Phi}{\partial \pmb{\sigma}}}\left(\bm{\sigma}_{n+1},p_{n+1},f_{n+1},\bm{\alpha}_n  \right) - (1-f_{n+1})\sigma_0(p_{n+1})\Delta p - \bm{\alpha}_n\gamma \Delta \Lambda \mathrm{tr}\left[\Phi\left(\bm{\sigma}_{n+1},p_{n+1},f_{n+1},\bm{\alpha}_n \right)\right] &= 0
  \end{aligned}
  \label{systeme}
\end{equation}
Non-linear system of equations (Eq.~\ref{systeme}) is solved using Newton-Raphson algorithm in the \texttt{MFront} code generator \cite{mfront}, where the jacobian matrix is evaluated numerically. Note that while the geometrical parameters (related to the void and cell aspect ratio $W$ and $\lambda$) are assumed to be fixed during the time step, transverse intervoid distance $\chi$ and interfacial strength are computed according to Eqs.~\ref{eqgamma} and \ref{eqchi}, respectively. The evolution of the geometrical parameters are computed explicitly according to Eqs.~\ref{evolporo},~\ref{evolgrowth},~\ref{evolcoa} and Appendix~A:
\begin{equation}
  \begin{aligned}
    \lambda_{n+1} &= \mathcal{F}\left(\lambda_n,\Delta \epsilon_p\right)\\
    W_{n+1} &= \mathcal{G}\left(W_n,f_n,\Gamma_n,\lambda_n,\Delta \epsilon_p \right)  
    \end{aligned}
\end{equation}
\section{Validation of the size-dependent homogenized model for isotropic  porous materials}
\subsection{Unit cell simulations}
\label{ucsimu}
\begin{figure}[H]
\centering
\includegraphics[height = 4.5cm]{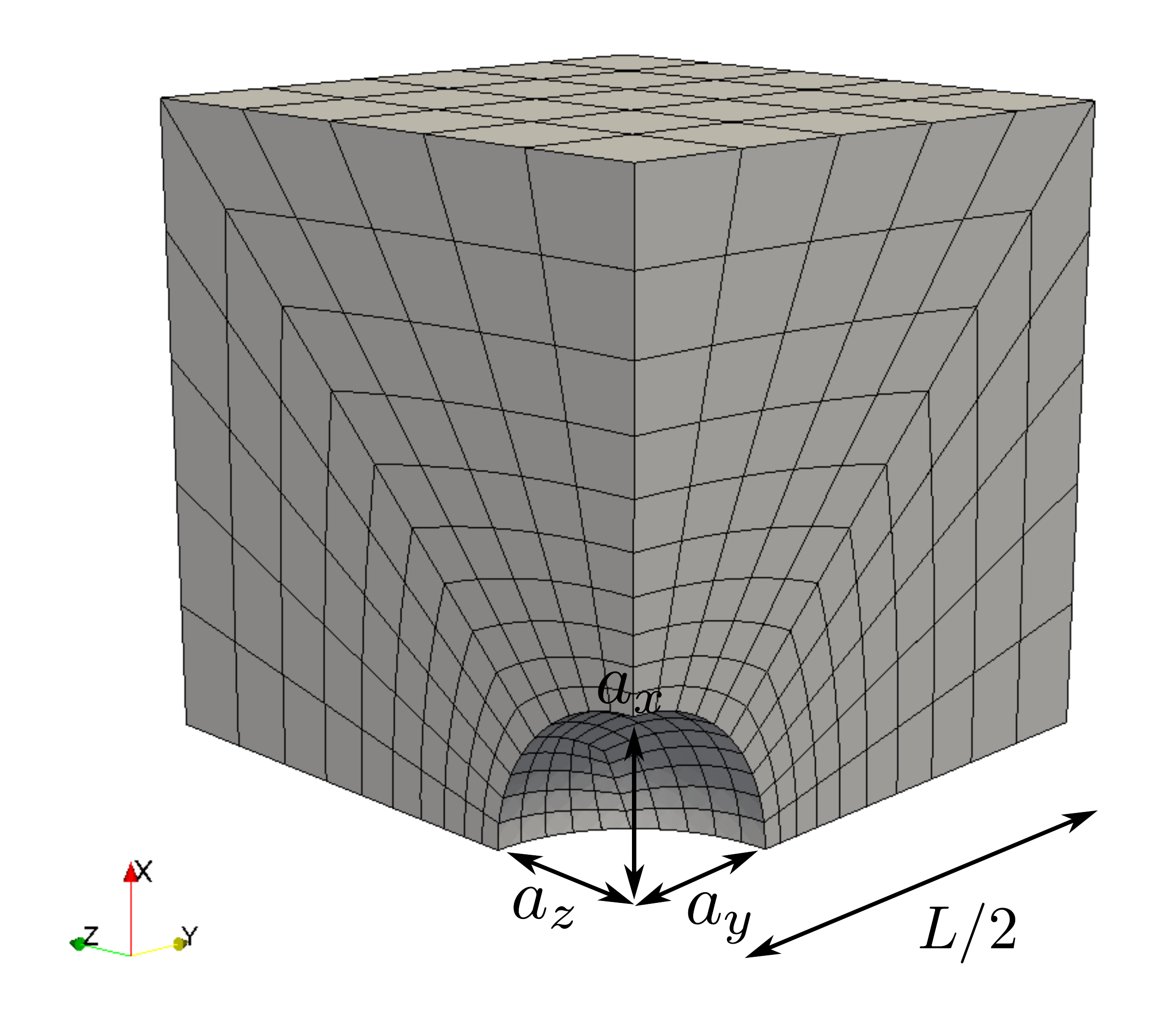}
\caption{Typical mesh used in finite element simulations, considering one-eighth of the unit cell}
\label{mesh}
\end{figure}

Finite strain unit cells simulations have been performed to validate the size-dependent homogenized model for porous materials described in Section~2. Porous cubic cells $\Omega_0$ of size $L$ under periodic boundary conditions with initially spherical voids of radius $R_0$ are considered (Fig.~\ref{mesh}), which corresponds to an initially simple cubic arrangement of spherical voids. Axisymmetric loading conditions are considered. Under such situation with an isotropic matrix material, external boundaries remain flat and parallel to each others, allowing to consider only one-eighth of the cell with the following boundary conditions:
\begin{equation}
  \begin{aligned}
    U_x(x=0,y,z) &= 0 \hspace{2cm}   U_x(x=L/2,y,z) &= U_x^{L/2} \\
    U_y(x,y=0,z) &= 0 \hspace{2cm}   U_y(x,y=L/2,z) &= U_{yz}^{L/2} \\
    U_z(x,y,z=0) &= 0 \hspace{2cm}   U_z(x,y,z=L/2) &= U_{yz}^{L/2} \\
  \end{aligned}
  \end{equation}
Macroscopic strain and stress of the unit-cell are computed such as:
\begin{equation}
    E_{xx} = \ln{\left(1 + \frac{2U_x^{L/2}}{L} \right)}\ \ \ \ \ \ \ \ \ \ E_{yy} = \ln{\left(1 + \frac{2U_{yz}^{L/2}}{L} \right)}\ \ \ \ \ \ \ \ \ \ E_{zz} = \ln{\left(1 + \frac{2U_{yz}^{L/2}}{L} \right)}
\end{equation}
\begin{equation}
  \bm{\Sigma} = \frac{1}{vol \Omega} \int_{\Omega} \bm{\sigma} d\Omega
\label{averagecauchy}
\end{equation}
where $\bm{\sigma}$ is the microscopic Cauchy stress, and $\Omega$ the deformed unit-cell. Constant macroscopic stress triaxiality $T$ are imposed:
  \begin{equation}
\bm{\Sigma} =
\left( \begin{array}{ccc}
\Sigma_{xx} & 0 & 0 \\
0 & \eta \Sigma_{xx} & 0 \\
0 & 0 & \eta \Sigma_{xx} \\
\end{array}
\right)  \ \ \ \ \  \ \ \ \ \  \Sigma_{eq} = \Sigma_{xx}(1 - \eta) \ \ \ \ \  \ \ \ \ \ \Sigma_m = \Sigma_{xx}\frac{1+2\eta}{3} \ \ \ \ \  \ \ \ \ \ T = \frac{\Sigma_{m}}{\Sigma_{eq}} = \frac{1+2\eta}{3(1-\eta)}
\label{axistress}
  \end{equation}
The displacement $U_x^{L/2}$ of the external boundary of normal $\underline{e}_x$ is imposed, and forces applied on the external boundaries of normal $\underline{e}_y$ and $\underline{e}_z$ are updated, for a given time step, during the iterations of the Newton-Raphson algorithm to solve global equilibrium according to $F_y = F_z = \eta \Sigma_{xx} S_{yz}$, where $S_{yz}$ is the updated surface of the boundaries, ensuring constant macroscopic stress triaxiality. Isotropic von Mises plasticity constitutive equations are used for the matrix material, with isotropic hardening according to:
\begin{equation}
\sigma_0(p)=\sigma_{0}\left(1+\frac{p}{p_0}\right )^m
\label{ecrouissage}
\end{equation}
while, consistently with the hypothesis used for the derivation of the yield criteria, 2D von Mises plasticity (of constant 2D plastic yield stress $\gamma$) is used for the void matrix interface. In practive, perfectly plastic (of yield stress $\sigma_0^{int}$)  shell elements are used, of thickness $t$ such that $\gamma = t\sigma_0^{int}$. Details about the modelling of the void matrix interface in finite element simulations and computation of the macroscopic stress through Eq.~\ref{averagecauchy} can be found in \cite{gallican}. 

\begin{figure}
	\subfigure{\includegraphics[trim=0.cm 24.cm 0.cm 0.cm, clip, width=.8\textwidth]{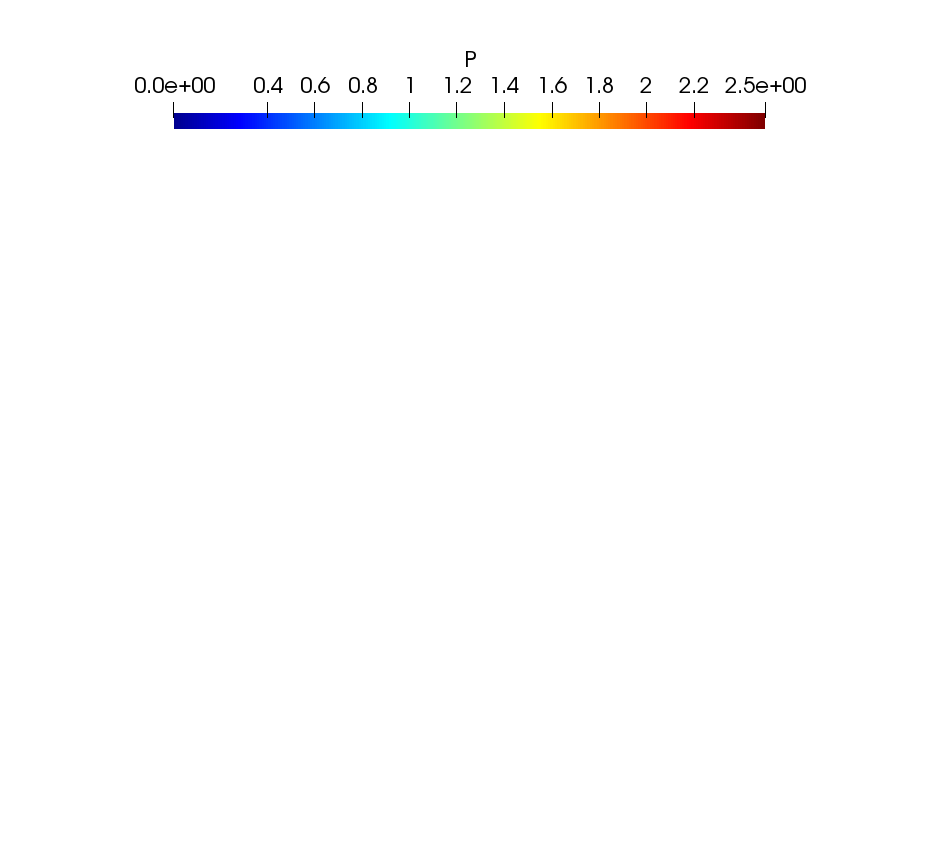}}
	\centering
	\setcounter{subfigure}{0}
	\subfigure[$\Gamma_0$=0]{\includegraphics[trim=1.cm 6.cm 1.cm 9.cm, clip, width=.25\textwidth]{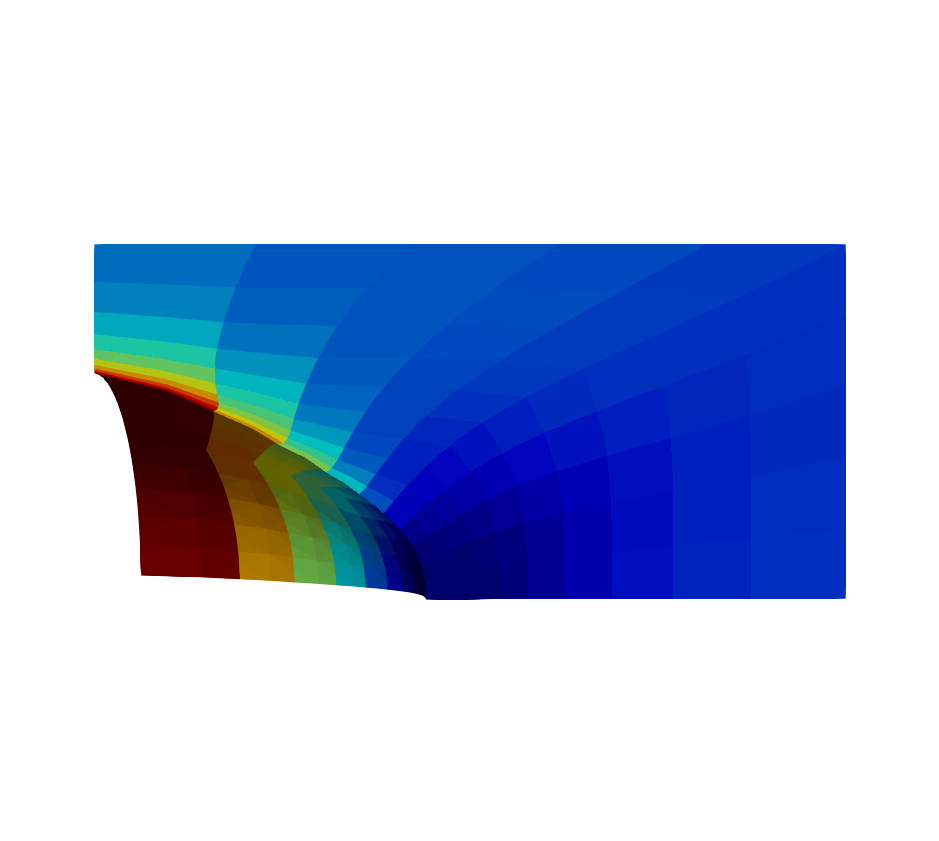}}
	\hspace{1cm}
	\subfigure[$\Gamma_0$=0.25]{\includegraphics[trim=1.cm 6.cm 1.cm 9.cm, clip, width=.25\textwidth]{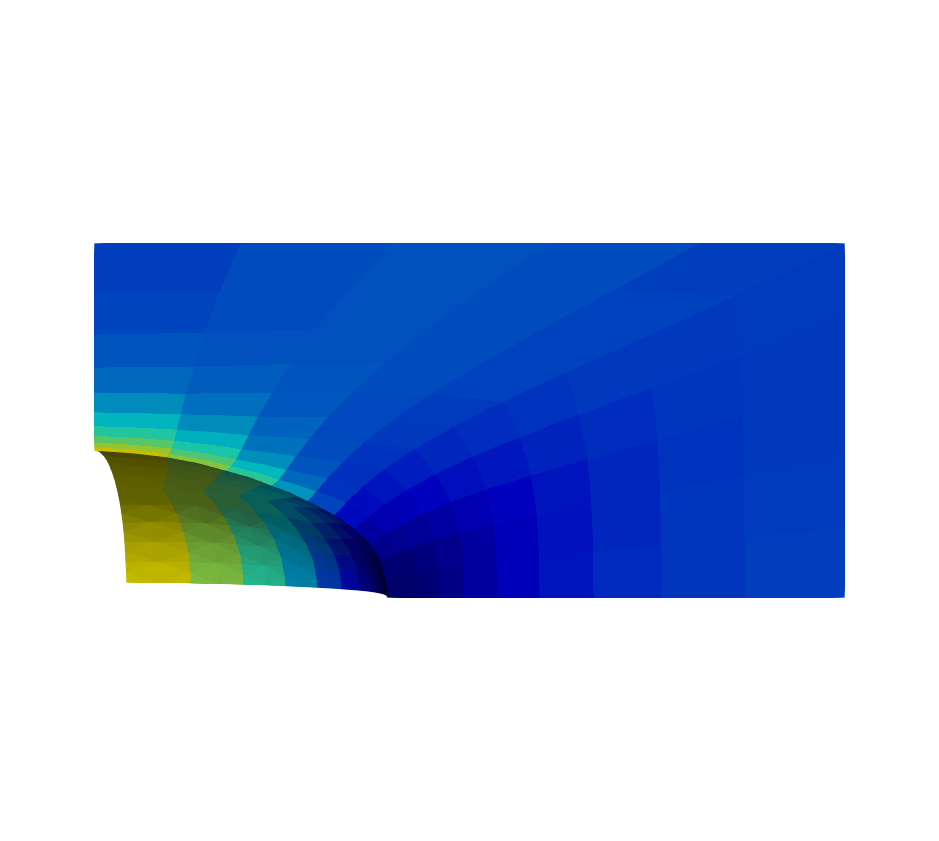}}\\
	\subfigure[$\Gamma_0$=0.5]{\includegraphics[trim=1.cm 6.cm 1.cm 7.cm, clip, width=.25\textwidth]{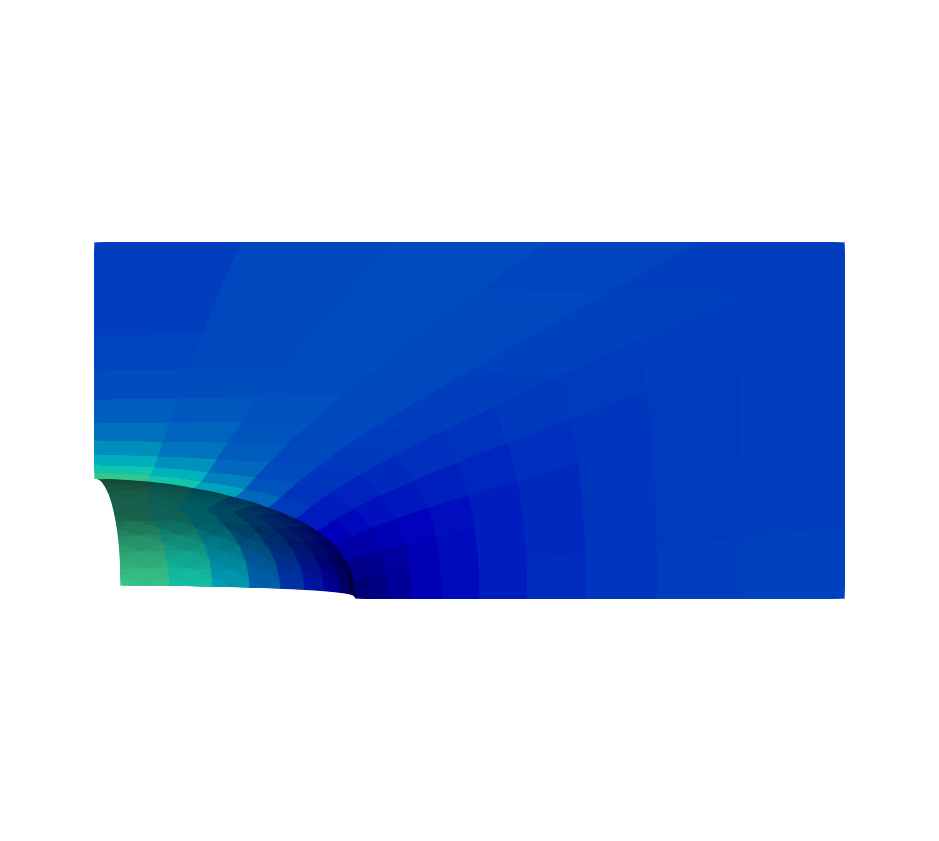}}
	\hspace{1cm}
	\subfigure[$\Gamma_0$=1]{\includegraphics[trim=1.cm 6.cm 1.cm 7.cm, clip, width=.25\textwidth]{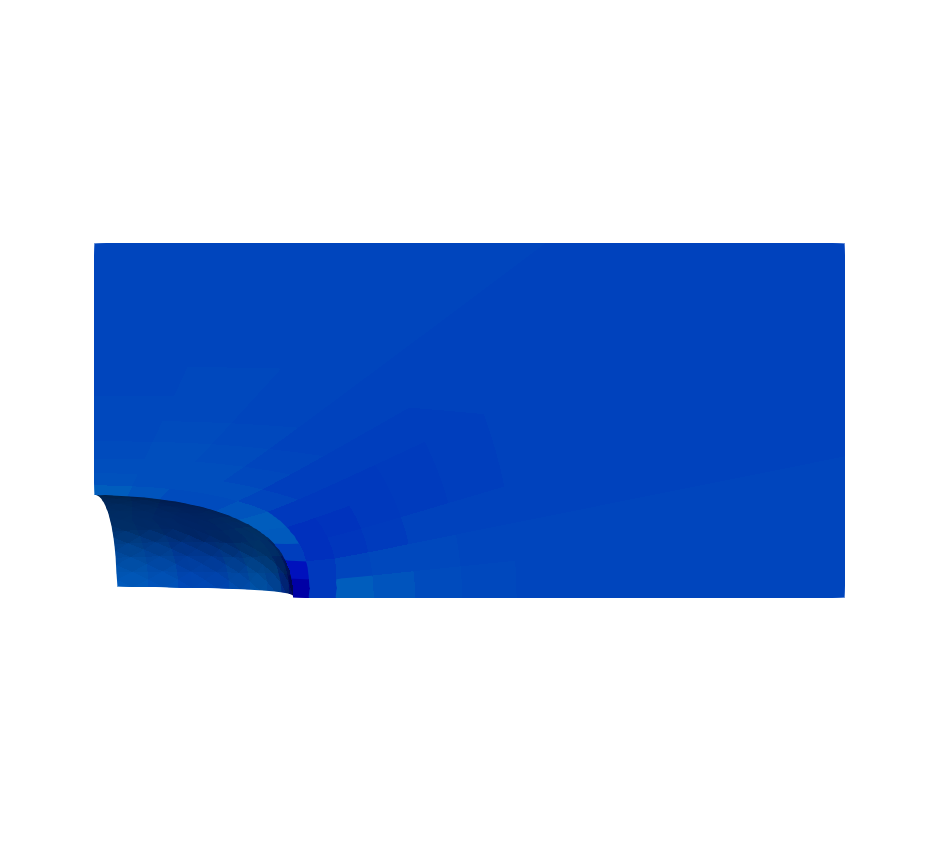}}
	\caption{Field of equivalent plastic strain for different initial interfacial strengths $\Gamma_0$ under axisymmetric loading conditions (the main loading direction corresponding to the horizontal axis), for an initial porosity of $f$=1\%, a stress triaxiality of $T=1$ and a macroscopic equivalent plastic strain of $E_{eq}=0.5$.}
	\label{fields}
\end{figure}

\begin{figure}
	\subfigure{\includegraphics[trim=0.cm 24.cm 0.cm 0.cm, clip, width=.8\textwidth]{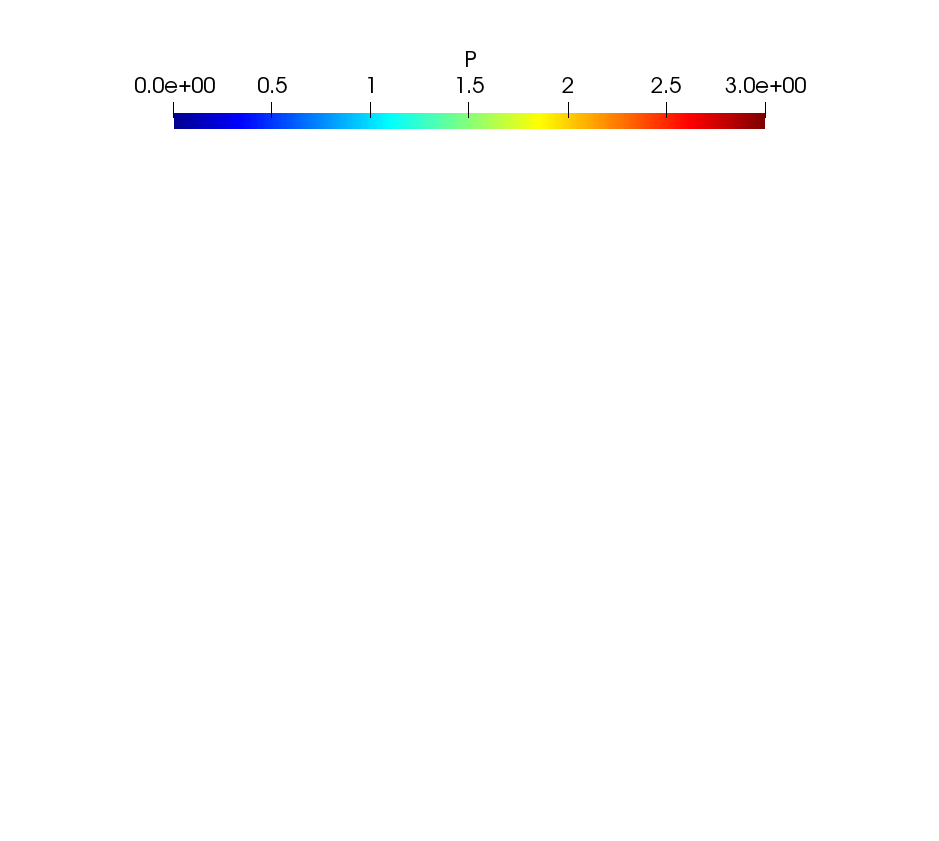}}
	\centering
	\setcounter{subfigure}{0}
	\subfigure[$\Gamma_0$=0]{\includegraphics[trim=1.cm 3.cm 1.cm 6.cm, clip, width=.25\textwidth]{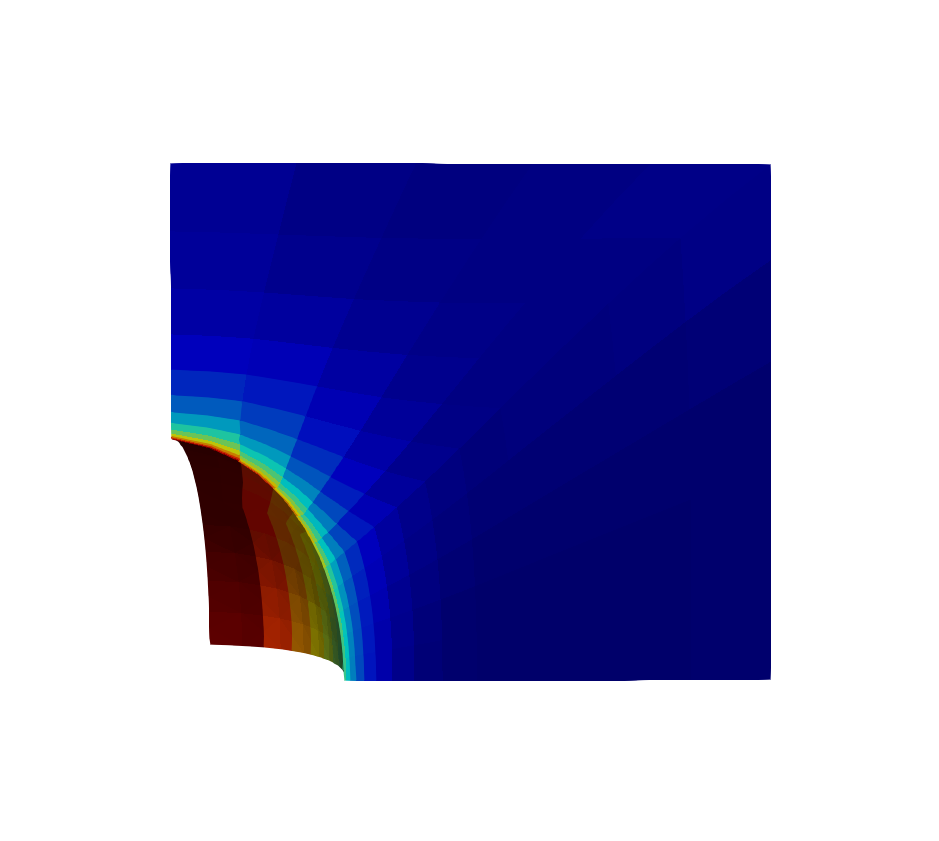}}
	\hspace{0.8cm}
	\subfigure[$\Gamma_0$=0.25]{\includegraphics[trim=1.cm 3.cm 1.cm 6.cm, clip, width=.25\textwidth]{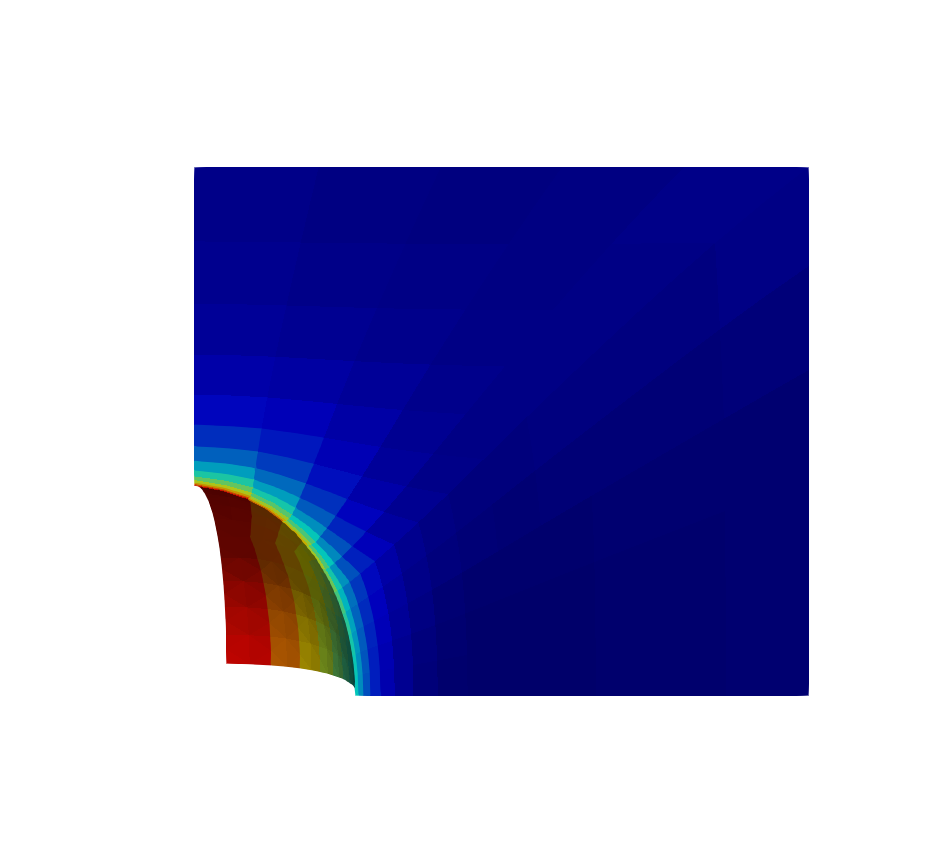}}\\
	\subfigure[$\Gamma_0$=0.5]{\includegraphics[trim=1.cm 3.cm 1.cm 6.cm, clip, width=.25\textwidth]{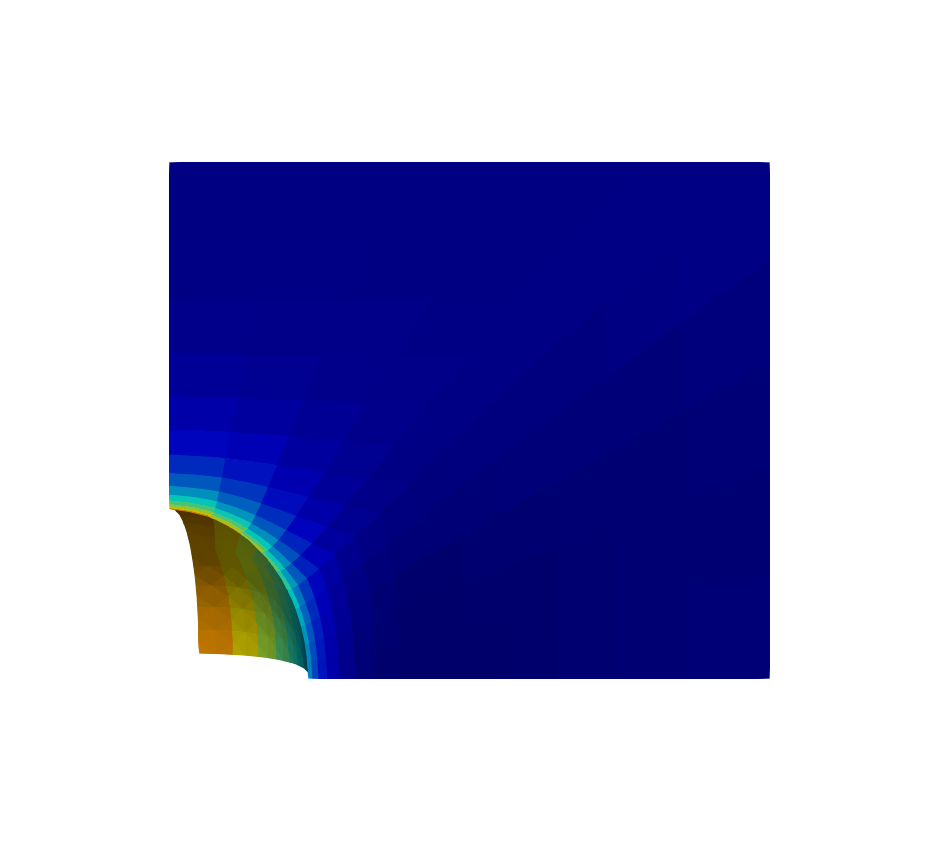}}
	\hspace{1cm}
	\subfigure[$\Gamma_0$=1]{\includegraphics[trim=1.cm 3.cm 1.cm 6.cm, clip, width=.25\textwidth]{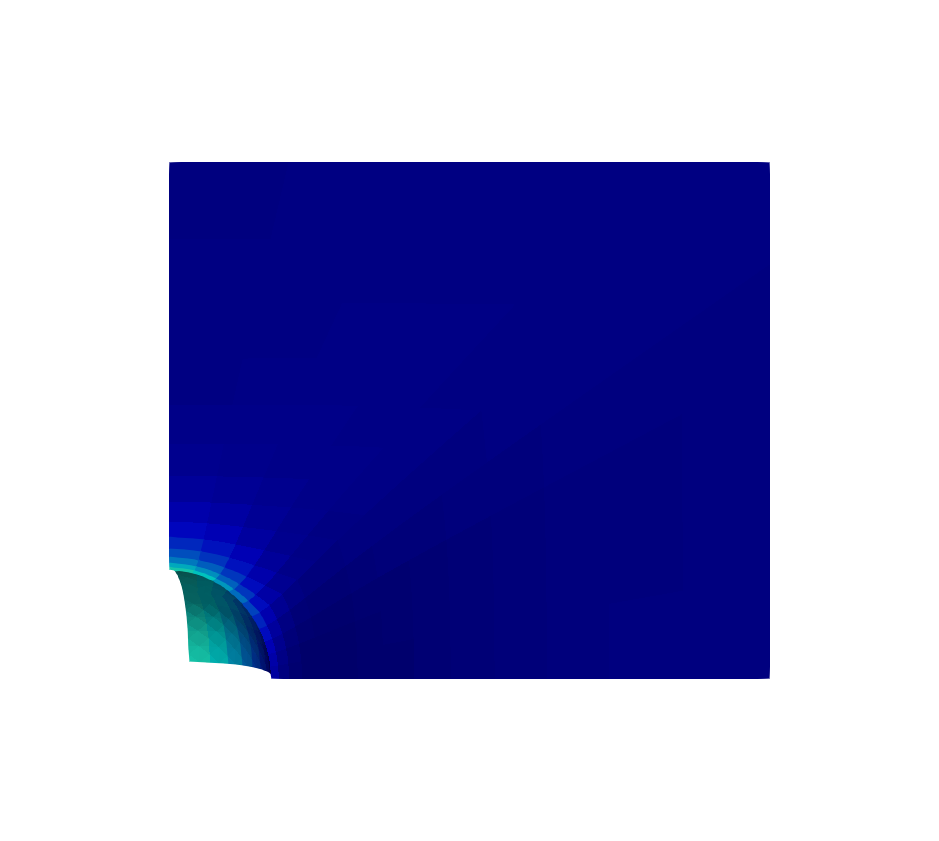}}
	\caption{Field of equivalent plastic strain for different initial interfacial strengths $\Gamma_0$ under axisymmetric loading conditions (the main loading direction corresponding to the horizontal axis), for an initial porosity of $f$=0.1\%, a stress triaxiality of $T=3$ and a macroscopic equivalent plastic strain of $E_{eq}=0.1$.}
	\label{fields2}
\end{figure}

\noindent 
Simulations have been performed with the finite-element solver \texttt{Cast3M} \cite{castem}. A typical mesh is shown on Fig.~\ref{mesh}, where quadratic brick elements and DKT elements are used for the matrix material and void matrix interface, respectively. Mesh convergence has been assessed for all results presented hereafter. Constitutive equations have been implemented in the \texttt{MFront} code generator \cite{mfront}, using the finite strain framework proposed in \cite{MAL}. The parameters assessed are the initial porosity $f_0$ and the initial interface strength $\Gamma_0$ defined as:
\begin{equation}
  f_0 = \frac{4\pi R_0^3}{3L^3} \ \ \ \ \ \ \ \ \ \ \Gamma_0 = \frac{\gamma}{\sigma_0 R_0}
\end{equation}
for a hardening exponent of $m=0.1$. For each simulation, the evolution of equivalent von Mises macroscopic stress, equivalent plastic strain, porosity and void aspect ratio are post-processed according to:
\begin{equation}
  \Sigma_{eq} = \sqrt{\frac{3}{2} \bm{\Sigma}':\bm{\Sigma}'} = \Sigma_{xx} - \Sigma_{yy} \ \ \ \ \ \ \ \ \ \ E_{eq} = \sqrt{\frac{2}{3}\textbf{E}':\textbf{E}'} = \frac{2}{3}|E_{xx} - E_{yy}| \ \ \ \ \ \ \ \ \ \ f = \frac{V_{void}}{V_{tot}} \ \ \ \ \ \ \ \ \ \ W = \frac{a_x}{a_y}
  \end{equation}
In Fig.~\ref{fields} equivalent plastic strain fields are presented for different values of $\Gamma_0$ for a stress triaxiality $T=1$ and at a macroscopic equivalent plastic strain field of $E_{eq}=50\%$. The stress triaxiality $T=1$ leads to prolate void shapes in all cases. The equivalent plastic strain field is more homogeneous for large values of $\Gamma_0$. Accordingly, the higher the value of $\Gamma_0$, the smaller the void deformation. In Fig.~\ref{fields2} equivalent plastic strain fields are presented for a higher stress triaxiality of $T=3$ at a macroscopic equivalent plastic strain field of $E_{eq}=10\%$. It can be observed that for all values of $\Gamma_0$ voids have an oblate shape. As $\Gamma_0$ increases, voids growth is more limited, and void shapes remain more spherical.
\color{black}

\subsection{Comparisons between homogenized model predictions and unit cell results}

The results of the homogenized model described in Section~\ref{consti} are compared to the results of the unit cell simulations (Section~\ref{ucsimu}) for axisymmetric loading conditions and various values of stress triaxialities $T \in [1;3]$, initial interfacial strength  $\Gamma_0 \in [0:1]$, initial porosity $f_0 \in [0.001;0.01]$ and for a hardening material $m=0.1$. An initial simple cubic lattice ($\lambda_0 = 1$) of spherical voids ($W_0=1$) is considered in all simulations. For the homogenized model, simulations are performed on a material point using the \texttt{MTest} software, and constant stress triaxiality is imposed through the use of Lagrange multipliers \cite{mfront}. The evolution law for the void aspect ratio has been calibrated and the final expression in given is Appendix~A. The values of the parameters $q_W$ (Eq.~\ref{qW}) and $q_{\chi}$ (Eq.~\ref{qchi}) are described hereafter.\\

\begin{figure}[H]
\centering
\subfigure[]{\includegraphics[height = 4.5cm]{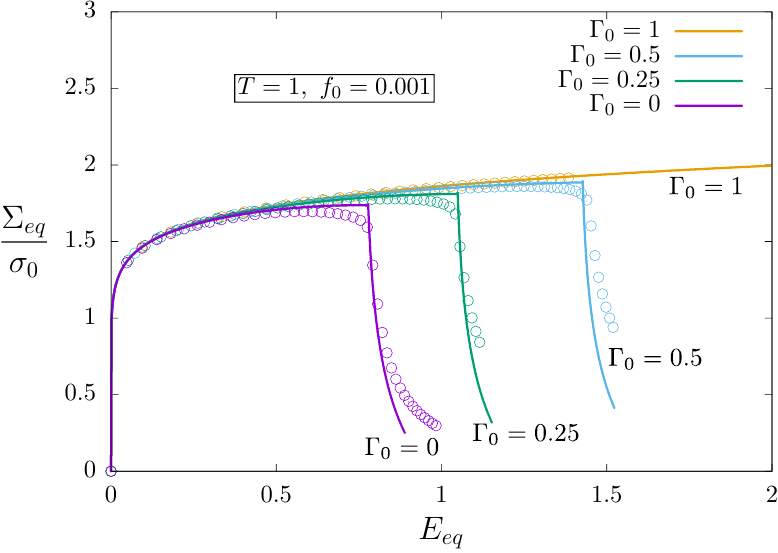}}
\hspace{1cm}
\subfigure[]{\includegraphics[height = 4.5cm]{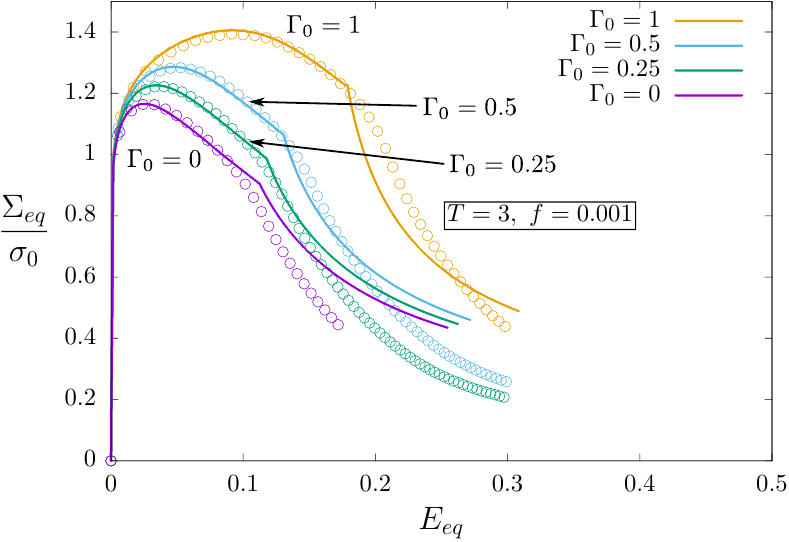}}
\caption{Normalized macroscopic von Mises stress as a function of macroscopic equivalent strain for different initial interfacial strengths $\Gamma_0$ under axisymmetric loading conditions, for an initial porosity of 0.1\% and for stress triaxiality of (a) $T=1$ and (b) $T=3$. Points correspond to the results of the unit cell simulations, lines to the homogenized model (with $q_{\chi} = 0.90$ for $T=1$, and $q_{\chi} = 0.62$ for $T=3$)}
\label{comp1}
\end{figure}

The evolution of the normalized macroscopic von Mises stress $\Sigma_{eq}/\sigma_0$ as a function of the macroscopic equivalent strain is given in Fig.~\ref{comp1}, for a stress triaxiality $T=1$ (Fig.~\ref{comp1}a) and $T=3$ (Fig.~\ref{comp1}b), for an initial porosity of 0.1\%. For low applied stress triaxiality $T=1$, a hardening behavior is observed for low applied strains, followed by a sudden softening corresponding to coalescence \cite{koplik}. As the initial dimensionless strength of the interface $\Gamma_0$ increases, two different effects are observed: a slight hardening at low applied strains, and more importantly a strong shift of the onset of coalescence towards higher strains. Similar observations can be made at higher stress triaxiality (Fig.~\ref{comp1}b) but with a less marked transition between void growth and void coalescence regimes.

\begin{figure}[H]
\centering
\subfigure[]{\includegraphics[height = 4.5cm]{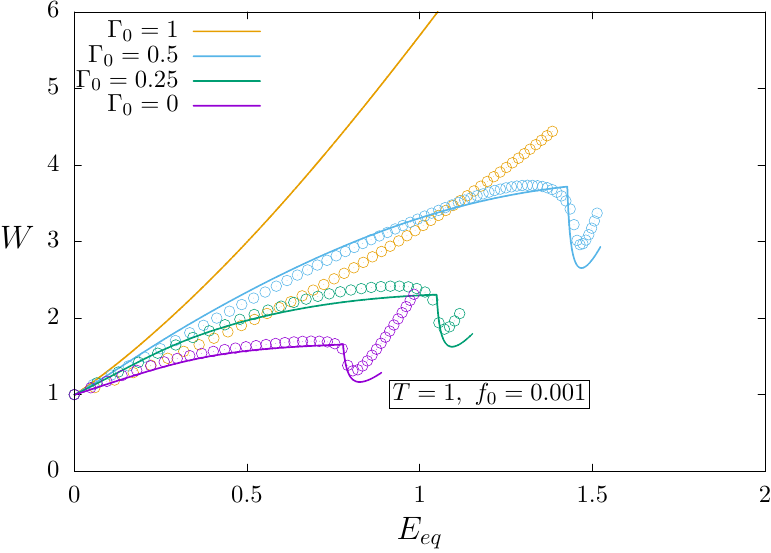}}
\hspace{1cm}
\subfigure[]{\includegraphics[height = 4.5cm]{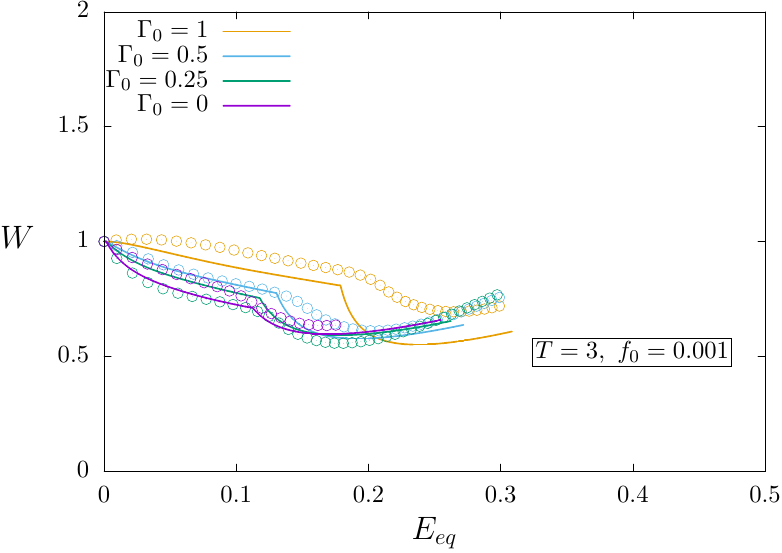}}\\
\subfigure[]{\includegraphics[height = 4.5cm]{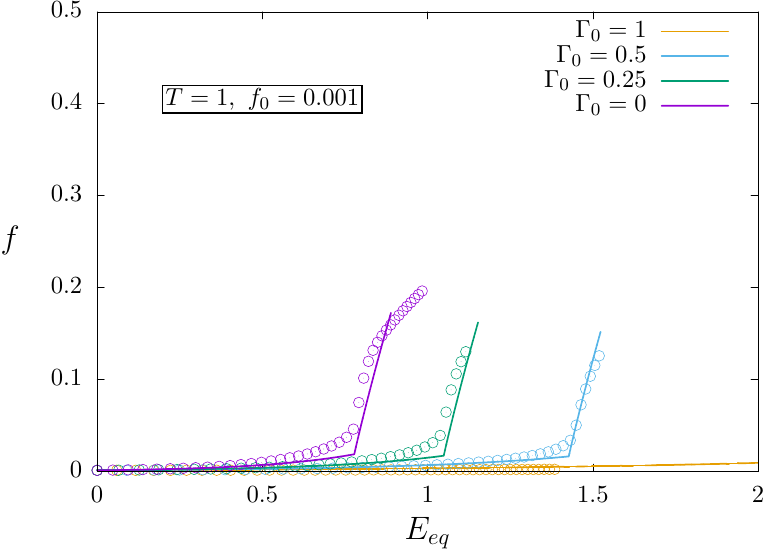}}
\hspace{1cm}
\subfigure[]{\includegraphics[height = 4.5cm]{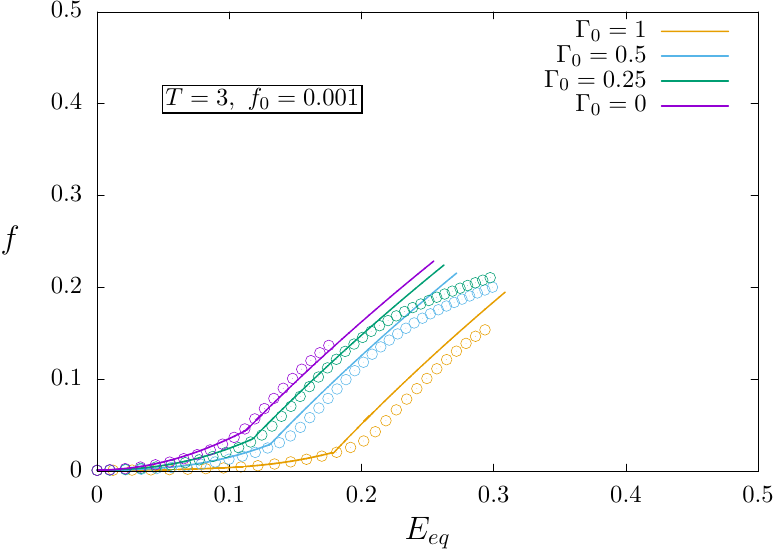}}
\caption{Void aspect ratio (a,b) and porosity (c,d) as a function of macroscopic equivalent strain for different initial interfacial strengths $\Gamma_0$ under axisymmetric loading conditions, for an initial porosity of 0.1\% and for stress triaxiality of $T=1$ and $T=3$. Points correspond to the results of the unit cell simulations, lines to the homogenized model (with $q_{\chi} = 0.90$ for $T=1$, and $q_{\chi} = 0.62$ for $T=3$)}
\label{comp2}
\end{figure}

The homogenized model is found to be in good agreement with unit cell simulations in the growth regime, \textit{i.e.} before the onset of coalescence, for both stress triaxialities and all values of dimensionless interfacial strength. As the yield criterion proposed in \cite{monchiet2013} was shown to be very accurate for the unit cell used in the derivation \cite{morinthese}, the agreement observed in Fig.~\ref{comp1} is rooted into the appropriate calibration of both the parameter $q_W$ added to the growth yield criterion, and to the evolution law for the evolution of the void aspect ratio $W$. For the former, the expression given in \cite{pardoen} has been used for $\Gamma = 0$, with a multiplicative factor related to $\Gamma$, as shown in Appendix~A. For the latter, the parameter $\mathcal{W}$ in Eq.~\ref{evolgrowth} has been calibrated based on the unit cell results (Fig.~\ref{comp2}a,b), leading to a very good agreement, except for high interfacial strength ($\Gamma_0 =1$), and the final expression is also given in Appendix~A. The onset of coalescence in unit cell results, corresponding to the change of slope in the evolution of macroscopic stress (Fig.~\ref{comp1}) related to the transition from void growth to void coalescence deformation mode, is well captured by the homogenized model, as well as the evolutions of stress and void aspect ratio in the coalescence regime. However, this agreement is obtained only through calibrating the parameter $q_{\chi}$ (Eq.~\ref{eqchi}) as a function of stress triaxiality,  with $q_{\chi} = 0.90$ for $T=1$, and $q_{\chi} = 0.62$ for $T=3$. Reminding that this parameter is introduced to use coalescence stresses derived from cylindrical voids and cylindrical unit cells for more realistic situation such as orthorhombic lattice of spheroidal voids, the dependence of $q_{\chi}$ to the stress triaxiality can be understood as follows. For $T=1$, an initially spherical void becomes prolate (Fig.~\ref{comp2}a), which is a situation where coalescence stress derived in \cite{torki,gallican} is very accurate, and $q_{\chi}$ accounts only for the difference of the unit cell. The value calibrated ($q_{\chi} = 0.90$) is close to the value used in \cite{barrioz2018,hure2018} for spheroidal voids in orthorhombic unit cells. For $T=3$, an initially spherical void becomes oblate, and coalescence stress derived in \cite{gallican} is known to be less accurate in that situation: $q_{\chi}$ accounts for both discrepancy of the coalescence yield criterion and difference of unit cell, hence the different calibrated value than for $T=1$. For all situations, the evolution of porosity (Fig.~\ref{comp2}c,d), deriving from mass conservation, is well captured by the homogenized model.
 
\begin{figure}[H]
\centering
\subfigure[]{\includegraphics[height = 4.5cm]{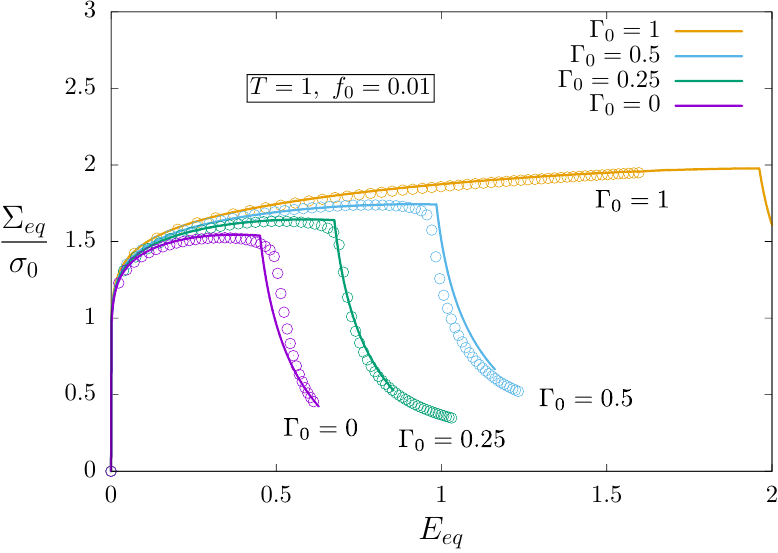}}
\hspace{1cm}
\subfigure[]{\includegraphics[height = 4.5cm]{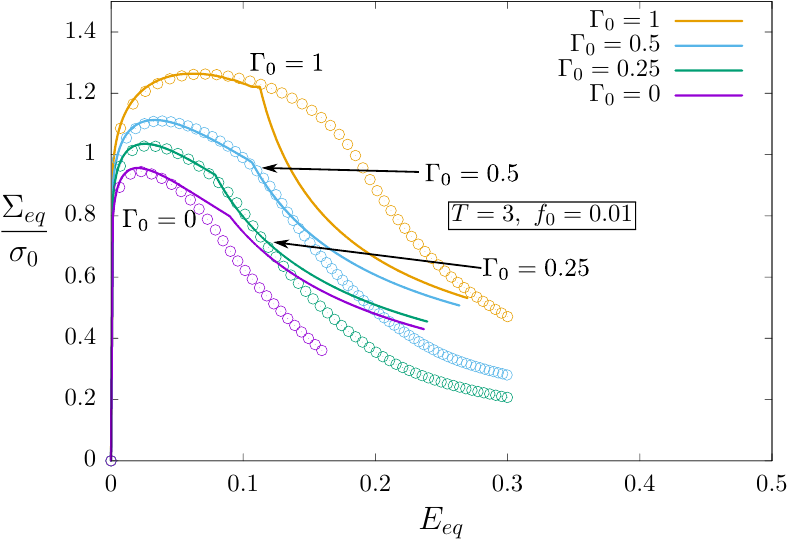}}\\
\subfigure[]{\includegraphics[height = 4.5cm]{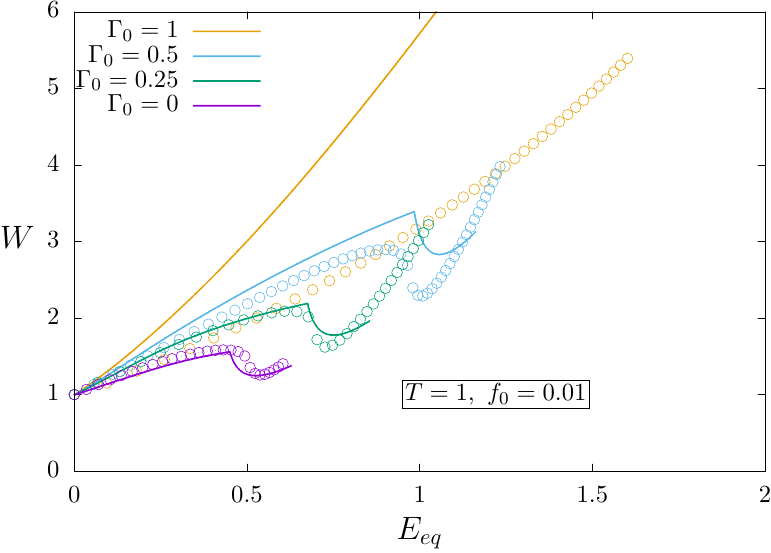}}
\hspace{1cm}
\subfigure[]{\includegraphics[height = 4.5cm]{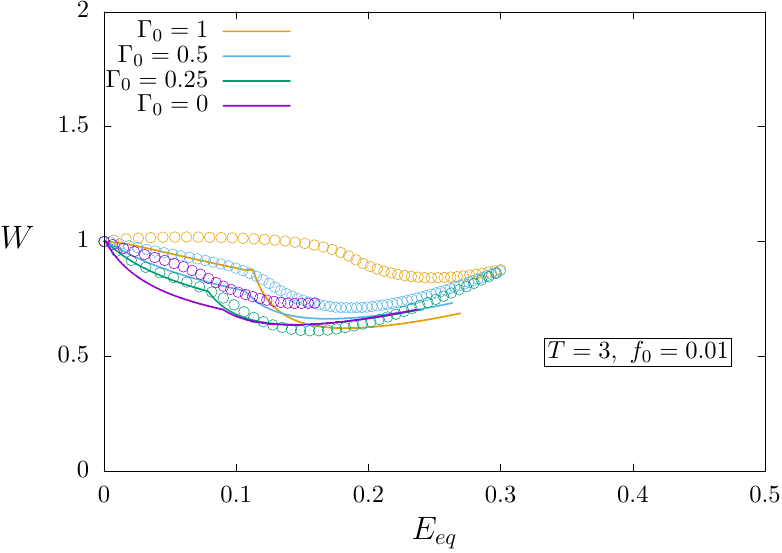}}\\
\subfigure[]{\includegraphics[height = 4.5cm]{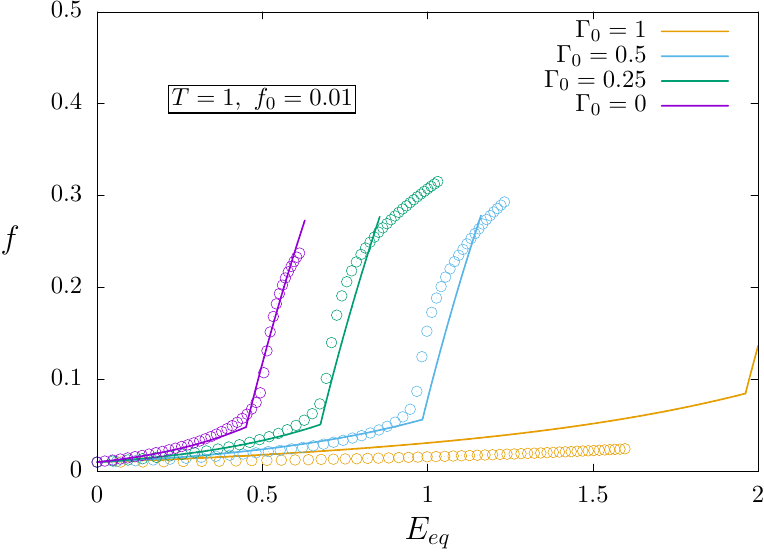}}
\hspace{1cm}
\subfigure[]{\includegraphics[height = 4.5cm]{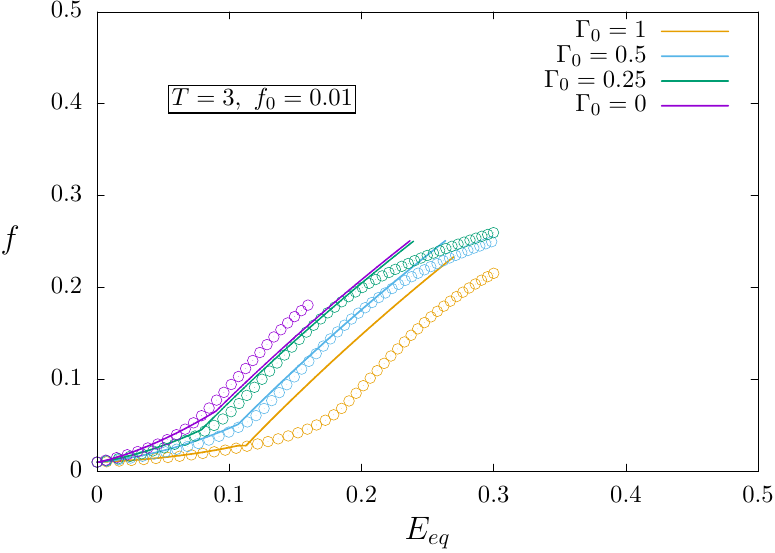}}
\caption{Normalized macroscopic von Mises stress (a,b), void aspect ratio (c,d) and porosity (e,f) as a function of macroscopic equivalent strain for different initial interfacial strengths $\Gamma_0$ under axisymmetric loading conditions, for an initial porosity of 1\% and for stress triaxiality of $T=1$ and $T=3$. Points correspond to the results of the unit cell simulations, lines to the homogenized model (with $q_{\chi} = 0.82$ for $T=1$, and $q_{\chi} = 0.62$ for $T=3$}
\label{backup}
\end{figure}

Comparisons are made between the results from unit cell simulations and the homogenized model for a larger value of the initial porosity ($f_0 = 1\%$) in Fig.~\ref{backup}. The evolutions of von Mises stress, void aspect ratio and porosity are found to be in good agreement in the growth regime, except for the highest interfacial strength. The quantitative predictions of the onset of coalescence still remains a challenge: the parameter $q_{\chi}$ calibrated as a function of stress triaxiality for the lower value of the initial porosity (Figs.~\ref{comp1},~\ref{comp2}) does not lead to quantitative agreement with the unit cell results. Fig.~\ref{backup} corresponds to $q_{\chi} = 0.82$ for $T=1$ (slightly lower that the one used for $f_0=0.1\%$), and $q_{\chi} = 0.62$ for $T=3$ (same value than for $f_0=0.1\%$). With these values, for $T=1$, a good agreement is observed in the coalescence regime regarding the evolutions of von Mises stress, void aspect ratio and porosity for the different initial interfacial strengths considered, while for $T=3$ the agreement is less quantitative. \\

\section{Discussion}

The comparisons detailed in Section~3 show that the size-dependent homogenized model for isotropic porous materials under axisymmetric loading conditions described in Section~2 leads to predictions in overall agreement with reference results obtained through porous unit cell simulations, and therefore can be used to model porous materials for which size effects are expected. \textcolor{black}{The range of validity of the model comes from the assumptions used in the derivation of the yield criteria, namely continuum mechanics and isotropic plastic flow. For nanometric voids, the first assumption seems justified in situations where dislocations (sources) density is high and/or large applied strain, as can be inferred from some experimental results \cite{margolin2016,ding2016}. The second assumption makes the model relevant for high stress triaxialities, where crystallographic-induced anisotropy has been shown to affect only weakly void deformation. Finally, although the equations of the model are presented in general terms, making them usable for arbitrary loading conditions, the yield criteria used are only valid for axisymmetric loading conditions, as discussed in Section~2.2.} Discrepancies between model predictions and reference results have been observed, requiring theoretical investigations to improve the current model. These investigations are outside the scope of this study, but potential research axis are described hereafter. The growth yield criterion used has been shown in \cite{monchiet2013} to be very accurate by comparison to numerical results obtained for the same geometry used in the derivation, \textit{i.e.}, confocal spheroidal void and cell \cite{morinthese}. Using this criterion for an initially simple cubic lattice of spherical voids, and including hardening, leads to results in good agreement with reference results, but requires calibrating a parameter $q_W$. A refined yield criterion is thus needed to prevent the use of this parameter, for example using non-linear variational homogenization as done for the development of models for porous materials without size effects \cite{danas}. This approach would also lead to theoretical expression for the evolution of void aspect ratio, removing the need for calibration. The modeling of void coalescence also requires improvements: most of the models predicting coalescence stress for porous materials developed recently \cite{torki,keralavarma,hurebarrioz} consider cylindrical voids in cylindrical unit cells, as the one used in this study \cite{gallican}, which poses some difficulties when applied to other configurations, such as orthorhombic lattice of spheroidal voids in this study. As proposed initially in \cite{torki}, a parameter $q_{\chi}$ may be introduced to go from one configuration to the other. However, a strong dependence of the onset of coalescence to this parameter was observed in this study. The effect of void distribution on coalescence stress should be investigated in more details, as for example initiated in \cite{hure2018}.

The application of the homogenized model described in this study is at first sight restricted to the case of nanoporous materials for which the contribution of the interface should be taken into account due to the presence of interface stresses / surface tension. The underlying physics correspond to the modelling of nanoporous materials initiated in \cite{dormieux2010} and lead to the yield criteria used in this study \cite{monchiet2013,gallican}. However, such models can be used to describe phenomenologically porous materials where size effects are expected through the proper calibration of the interfacial strength $\Gamma$. Size effects for porous materials are expected from numerical simulations due to the presence of Geometrically Necessary Dislocations and modelled through strain-gradient plasticity \cite{borg2008}. Typical porous unit cell simulations results are for example provided in \cite{niordson2008} where it is observed higher yield stress and delayed coalescence as the characteristic lengthscale introduced by the strain-gradient model decreases, in a very similar way as what is observed in this study considering interfacial dissipation. Therefore, the dimensionless interfacial strength $\Gamma$ could be considered as a parameter representing, in a simplified way, the additional hardening occurring close to the void matrix interface due to the presence of GNDs. It is expected that a proper calibration of $\Gamma$ with respect to the characteristic lengthscale used in strain-gradient model should lead to predictions in good agreement with the results shown in \cite{niordson2008}. 

\newpage
\section{Conclusion}
\color{black}
A size dependent homogenized model for isotropic porous materials is described in this study based on yield criteria derived for nanoporous material in growth \cite{monchiet2013} and coalescence \cite{gallican} regimes, and adding evolutions laws for the hardening, porosity, cell and void aspect ratios. The latter has been calibrated through comparisons to reference finite strain porous unit cell simulations incorporating interfacial stresses. A good agreement between the predictions of the homogenized model and the reference simulations regarding stress, porosity and void shape is obtained under axisymmetric loading conditions over a large range of interfacial strength. Two phenomenological parameters, denoted $q_{W}$ and $q_{\chi}$ and classically used to improve the predictions of porous material yield criteria, have been also calibrated but should be the subject of further studies. In particular, a rather strong dependence of the predictions to the parameter $q_{\chi}$ has been observed, and deeper investigation is therefore necessary to better assess the effect of void distribution on coalescence. The homogenized model can be used to describe porous materials once calibrating the interfacial strength, either based on lower scale simulations or directly from experiments which are still lacking in the literature and deserve more attention.\\

\noindent
\textbf{Acknowledgements}\\

The authors would like to thank L{\'e}o Morin and Jacques Besson for fruitful discussions.

\color{black}
\newpage

\section{Appendix A}
\label{appendix1}

The geometrical parameters involved in the growth yield criterion (Eq.~\ref{critgrowth1}) are detailed in Tab.~\ref{tabgrowth}.
\begin{table}[H]
  \scalebox{0.8}{
	\centering
	\caption{Geometrical parameters involved in Eq.~\ref{critgrowth1}, taken from \cite{benzergaleblond,monchiet2013}}
	\label{parametres_spheroides}
	\renewcommand{\arraystretch}{1.6}
	\begin{tabular}{|p{4.cm}|c|c|}
		\hline
		\centering   & Prolate void & Oblate void \\ \hline
		\centering Void aspect ratio $W$ & \multicolumn{2}{|c|}{$\displaystyle{\frac{a_1}{b_1}}$}\\ \hline
		\centering Porosity $f$ & \multicolumn{2}{|c|}{$\displaystyle{{a_1b_1^2}/{a_2b_2^2}}$}\\ \hline
		\centering Focal length $c$ & $\sqrt{a_1^2-b_1^2}=\sqrt{a_2^2-b_2^2}$ & $\sqrt{b_1^2-a_1^2}=\sqrt{b_2^2-a_2^2}$\\ \hline
		\multirow{2}{*}{ \parbox{1\linewidth}{\hspace{0.4cm} Void excentricity $e_1$}  }& $\displaystyle{\frac{c}{a_1}}$ & $\displaystyle{\frac{c}{b_1}}$\\ \cline{2-3}
		& \multicolumn{2}{|c|}{$\sqrt{1-\min{\left(W^2,\frac{1}{W^2}\right)}}$} \\ \hline
		\multirow{2}{*}{ \parbox{1\linewidth}{\hspace{0.4cm} Cell excentricity $e_2$} } & $\displaystyle{\frac{c}{a_2}}$ & $\displaystyle{\frac{c}{b_2}}$\\ \cline{2-3}
		& \multicolumn{2}{|c|}{$\displaystyle{\frac{e_2^3}{\sqrt{1-e_2^2}}=f\frac{e_1^3}{\sqrt{1-e_1^2}} }$} \\ \hline \hline
		\centering $\alpha$ & $ \displaystyle{\frac{1-e^2}{e^3}(\tanh^{-1}(e)-e)}$ & $\displaystyle{\frac{e-\sin^{-1}(e)\sqrt{1-e^2}}{e^3}}$ \\ \hline
                		\centering $g\ \ / \ \ g_i$ & \multicolumn{2}{|c|}{$\displaystyle{\frac{e_2^3}{\sqrt{1-e_2^2}}\ \ / \ \  \frac{g}{g + i}}$}\\ \hline
		\centering $\kappa$ & $\displaystyle{\left[\frac{1}{\sqrt{3}} + \frac{1}{\ln f}\left((\sqrt{3}-2) \ln{\frac{e_1}{e_2}}   \right)    \right]^{-1}}$ & $\displaystyle{\frac{3}{2} \left[1 +  \frac{(g_f - g_1) + \frac{4}{5}(g_f^{5/2} - g_1^{5/2}) - \frac{3}{5}(g_f^{5} - g_1^{5})  }{ \ln{\frac{g_f}{g_1}}}    \right]^{-1}}$\\ \hline
		\centering $\displaystyle{\frac{S_1}{V_1}}$ &  \multicolumn{2}{|c|}{$\displaystyle{\frac{3}{2a_1}\left(\frac{W}{\sqrt{\left|1-\frac{1}{W^2}\right|}}\tan^{-1}\left(\sqrt{\left|W^2-1\right|}\right)+1\right)}$}  \\ \hline
		\centering $u_1 \ \ / \ \ u_2$ & \multicolumn{2}{|c|}{$\displaystyle{\frac{f}{1+g} \ \ / \ \ \frac{f}{f+g}}$} \\ \hline
		\centering $\eta$ & \multicolumn{2}{|c|}{$\displaystyle{\eta=\frac{\kappa^2(1+g)(f+g)(\alpha_2-\alpha_1)}{(1-f)}}$}\\ \hline
		\centering $\zeta$ & \multicolumn{2}{|c|}{$\displaystyle{\zeta=\frac{\kappa^2(1+g)(f+g)(\alpha_2-\alpha_1)^2}{(1-f)^2}}$}\\ \hline
		\centering $\gamma$ & \multicolumn{2}{|c|}{$\displaystyle{\frac{1}{W^2 - 1} \left(1 - \frac{3V_1}{aS_1}    \right)}$}\\ \hline
		\centering $\mu$ & \multicolumn{2}{|c|}{$\displaystyle{\frac{W^2}{W^2-1}(1-3\gamma)}$}\\ \hline
		\centering $U$ & \multicolumn{2}{|c|}{$\displaystyle{1-3\alpha_1-f(1-3\alpha_2)}$}\\ \hline
		\centering $h_1$ &\multicolumn{2}{|c|}{ $\displaystyle{(3\gamma+3\mu-2)U^2+2(3\gamma-1)U+4}$} \\ \hline
		\centering $h_2$ &\multicolumn{2}{|c|}{$\displaystyle{f^2(3\gamma+3\mu-2)}$} \\ \hline
		\centering $h_3$ &\multicolumn{2}{|c|}{$\displaystyle{f(3\gamma+3\mu-2)U+f(3\gamma-1)}$} \\ \hline
                \centering $\mathcal{X}(\xi)$    &\multicolumn{2}{|c|}{$\displaystyle{\sqrt{(1-\zeta)(\xi + (1-3\alpha_2))^2}}$} \\ \hline
                \centering  $\mathcal{Y}(\xi)$   &\multicolumn{2}{|c|}{$\displaystyle{[{3 + \eta(1-3\alpha_2) + \eta \xi}]/{f \kappa}}$} \\ \hline
                \centering $\mathcal{Z}(\xi)$    &\multicolumn{2}{|c|}{$\displaystyle{\sqrt{h_1 + h_2\xi^2 + 2h_3}}$} \\ \hline
                \centering  $\mathcal{U}(\xi)$   &\multicolumn{2}{|c|}{$\displaystyle{f\left[\mathrm{sinh^{-1}}\left(\frac{u \mathcal{Y}(\xi)}{\mathcal{X}(\xi)}   \right)    \right]_{u_1}^{u_2}}$} \\ \hline
                \centering  $\mathcal{V}(\xi)$   &\multicolumn{2}{|c|}{$\displaystyle{f\left[\frac{\sqrt{\mathcal{X}^2(\xi) + u^2\mathcal{Y}^2(\xi)}}{u \mathcal{X}^2(\xi)}    \right]_{u_1}^{u_2}}$} \\ \hline
       \centering$\displaystyle\beta_1$& $\left[{e_1 - (1-e_1^2)\mathrm{tanh}^{-1}(e_1)}\right]/{2e_1^3}$& $\left[{-e_1(1-e_1^2) + \sqrt{1-e_1^2}\mathrm{sin}^{-1}(e_1)}\right]/{2e_1^3}$\\ \hline
       \centering$\displaystyle\beta_2$& $(1+e_2^2)/(3+e_2^4)$& $(1-e_2^2)(1-2e_2^2)/(3-6e_2^2+4e_2^4)$\\ \hline
	\end{tabular}
        \label{tabgrowth}
        }
\end{table}

\section{Appendix B}

The parameter $q_W$ involved in the definition of the effective porosity in the growth yield criterion (Eq.~\ref{critgrowth1}) is taken from \cite{pardoen} for $\Gamma = 0$, with a multiplicative term to account for the interfacial strength:
\begin{equation}
  q_W = (1+0.47\Gamma)\left[\mathrm{tan}^{-1}(4(2.5 - T))\frac{|b-1|}{\pi} + \frac{b+1}{2}\right]
\end{equation}
where $T$ is the stress triaxiality, and $b = 1 + (0.655-1.75m-0.544\sqrt[4]{f})\left(0.5 + [\mathrm{atan}(2(1-S))]/\pi - 0.0288\exp{(-1.08(0.2+S)}    \right)$, $f$ the porosity, $m$ the hardening exponent and $S=\ln W$ with $W$ the void aspect ratio.\\

\noindent
The heuristic correction for the evolution of the void aspect ratio in the growth regime has been calibrated against comparisons to unit cell results, and the final form is:
\begin{equation}
\mathcal{W} = \left[\sqrt{1.85\Gamma}(1.85-T^{0.87}+1)     \right] \left[ 1 - 2(1 - \sqrt{\Gamma})^3 \frac{\tan^{-1}{(2\ln{W})^4}}{\pi}   \right]    [2 - T^{2.15} + (T^2 - 1)e_1]
\label{eqW}
\end{equation}
In particular, for $\Gamma=0$ and spherical voids $W=1$, Eq.~\ref{eqW} reduces to:
\color{black}
\begin{equation}
  \mathcal{W} = 2 - T^{2.15}
  \label{eqW2}
\end{equation}
which is close to the original expression proposed in \cite{gologanu1993}.\\
\color{black}


\bibliographystyle{elsarticle-num.bst}
\bibliography{spebib2}

\end{document}